\documentclass[twocolumn]{aastex63}
\received{2020 August}
\revised{2020 September 23}
\accepted{2020 October 21}
\submitjournal{APJ}

\shorttitle{AGN ghosts}
\shortauthors{Esparza-Arredondo et al.}
\graphicspath{{./}{figures/}}

\begin{document}
\title{Active galactic nuclei ghosts: A systematic search for faded nuclei}

\correspondingauthor{Donaji Esparza-Arredondo}
\email{d.esparza@irya.unam.mx}

\author[0000-0001-8042-9867]{Donaji Esparza-Arredondo}
\affiliation{Instituto de Radioastronomía and Astrofísica (IRyA-UNAM)\\
3-72 (Xangari), 8701\\
Morelia, México}

\author[0000-0002-3467-8077]{Natalia Osorio-Clavijo}
\affiliation{Instituto de Radioastronomía and Astrofísica (IRyA-UNAM)\\
3-72 (Xangari), 8701\\
Morelia, México}

\author[0000-0002-2356-8358]{Omaira González-Martín}
\affiliation{Instituto de Radioastronomía and Astrofísica (IRyA-UNAM)\\
3-72 (Xangari), 8701\\
Morelia, México}

\author{Cesar Victoria-Ceballos}
\affiliation{Instituto de Radioastronomía and Astrofísica (IRyA-UNAM)\\
3-72 (Xangari), 8701\\
Morelia, México}

\author[0000-0001-9385-4176]{Sinhué Haro-Corzo}
\affiliation{Escuela Nacional de Estudios Superiores (ENES-UNAM)\\
3-72 (Xangari), 8701 \\
Morelia, México}

\author[0000-0001-7707-7389]{Ulises Reyes-Amador}
\affiliation{Instituto de Radioastronomía and Astrofísica (IRyA-UNAM)\\
3-72 (Xangari), 8701\\
Morelia, México}

\author{Jafet López-Sánchez}
\affiliation{Escuela Nacional de Estudios Superiores (ENES-UNAM)\\
3-72 (Xangari), 8701 \\
Morelia, México}

\author[0000-0003-1933-4636]{Alice Pasetto}
\affiliation{Instituto de Radioastronomía and Astrofísica (IRyA-UNAM)\\
3-72 (Xangari), 8701\\
Morelia, México}

\begin{abstract}
Physical processes such as re-ignition, enhancement, and fading of active galactic nuclei (AGN) are not entirely understood because the timeline of these events is expected to last many years. However, it is well known that the differences in the energy budget between AGN components, like the optical ionizing region and the mid-infrared (MIR) dust echoes, can be interpreted as a hint on AGN evolution. Here we present a catalog of 88 AGN candidates showing hints on the fading and rising of their activity in the nearby Universe. We use AGN scaling relations to select them from an initial sample of 877 candidates using publicly available optical, X-ray, and MIR luminosities. We then use the multi-wavelength information to discard sources contaminated with extranuclear emission and those with an X-ray luminosity not well corrected for absorption. We find that 96\% of our candidates are fading sources. This result suggests a scenario where the Universe had its peak of AGN activity somewhere in the past and is dominated by a fading phase at the present time. Alternatively, the fading phase is longer than the rising phase, which is consistent with galaxy merger simulations. Around 50\% of these fading candidates are associated with merging or interacting systems. Finally, we also find the existence of jets in $\rm{\sim}$30\% of these candidates and that the preferred AGN dust geometry is torus-like, instead of wind-like. Our results are compatible with the fading of nuclear activity, expected if they are in an inefficient state.
\end{abstract}



\section{Introduction}\label{intro}

Some of the most important questions in the active galactic nuclei (AGN) field are how and why AGN initiate or finish their activity. Understanding this behaviour plays a key role in the context of the supermassive black hole (SMBH) growth, which is linked to these active phases, and the evolution of the galaxies itself \citep[see ][]{Hopkins10}. Indeed, it is well known that the mass of the SMBH is linked to other properties of galaxies \citep[e.g.][]{Kormendy13}.

Despite its importance, little is known about it, with only a brief idea of the AGN ignition/fading process and/or the duration of the AGN phase. \citet{Marconi04} suggested that this phase should last $\rm{10^{7-9}}$ years spread in small duty cycles of $\rm{10^{5}}$ years each \citep[see also][]{Novak11,Schawinski15,Shulevski15}. Under this context, the study of AGN duty cycle cannot be done without a proper classification of the stages of the AGN. 

It has long been known that some AGN are accompanied by emission-line regions both narrow and broad (the so-called NLR and BLR). The NLR is a zone of ionized gas spanning galaxy scales or even larger. Such regions can trace the geometry of the ionizing radiation escaping from the AGN and the host galaxy, and at least implicitly can give hints to the AGN luminosity when this structure was created \citep{Keel17}. Indeed, the difference in the energy budget between the accretion disk and the NLR can be interpreted as a hint on the AGN evolution. This is what is called the {\emph {optical ionization echoes}}. In this way several fading AGN have been discovered by the Galaxy Zoo project \citep[][]{Lintott08}. A very well known example of a fading AGN discovered using this method is the Hanny’s Voorwerp near the spiral galaxy IC\,2497 \citep[][]{Lintott09,Keel12}. This object shows a NLR spanning a projected range from 15-35 kpc from the galaxy nucleus, that should have been produced by an AGN at least two orders of magnitude in bolometric luminosity higher than the nuclear luminosity. This indicates that the nucleus faded from a QSO-like luminosity to a modest Seyfert/LINER level within $\rm{10^{5}}$ years. AGN showing this scenario have also been reported in the high redshift Universe by the discovery of 14 Lyman-$\rm{alpha}$ blobs with weak AGN activity \citep[e.g.][]{Schirmer16}.

Extrapolating this line of thought, not only the NLR can trace these {\emph {echoes}} of past activity but also other components could help to find changes on the AGN activity. Mid-infrared (MIR) wavelengths can be used to trace the {\emph {MIR dust echoes}} because this emission is dominated by the obscuring dust located few parsecs away from the nucleus \citep[][]{RamosAlmeida17}. Of course, the closer the structure to the accretion disk, the shorter is the timescale of the evolution. Thus, in a fading scenario, we would expect the bolometric luminosity required for the NLR to be higher than that required for the AGN dust, with the accretion disk current bolometric luminosity being the lowest among them. This idea has already been applied to the case of Arp\,187 showing a clear decline of the nuclear activity (with over 10$^3$ times lower luminosity) in an estimated lapse of time of 10$^4$~years \citep[][]{Ichikawa16,Ichikawa18,Ichikawa19}. \citet{Ichikawa19} claim that the nucleus of Arp\,187 has already ceased its activity, with its NLR and jet being the evidence of the past activity. Despite its importance very few fading AGN have been reported so far. The largest compilation of them shows $\rm{\sim}$20 AGN \citep[see Table 2 presented by][and references therein]{Ichikawa19}. 
This technique could in principle not only detect fading AGN but also rising AGN, i.e., those AGN that show an increased accretion disk bolometric luminosity compared to the {\emph {MIR dust echoes}} and the {\emph {optical ionization echoes}}. Currently, the detection of rising-AGN candidates using only the NLR emissions is difficult because they may be indistinguishable from lack of gas for ionization echoes \citep[][]{Schawinski15}. In general terms, fading (rising) AGN show large-scale signatures of a prominent (weak) AGN process and small-scale signatures of a much weaker (stronger) AGN.

The purpose of this paper is to compile a sample of fading and rising AGN candidates using multi-wavelength information. The paper is organised as follows. In  Section\,\ref{sec:scalingrelations}, we use individual scaling relations found in AGN to derive an initial sample of candidates using publicly available optical and MIR observations of AGN. This sample is further refined in Section\,\ref{sec:candidates} combining together accurate measurements of the disk, AGN dust, and NLR for the initial sample. In Section\,\ref{sec:robustness}, we further explore the robustness of the selection of candidates using all available multi-wavelength information. We give a summary and discuss the main results in Section\,\ref{sec:discussion}. Finally, main conclusions of the paper are presented in Section\,\ref{sec:conclusions}. Throughout this work we adopt H$_0=67.8$ km~s$^{-1}$~Mpc$^{-1}$ as cosmological parameter.

\begin{figure*}
 \centering
 \includegraphics[width = 1.0\linewidth]{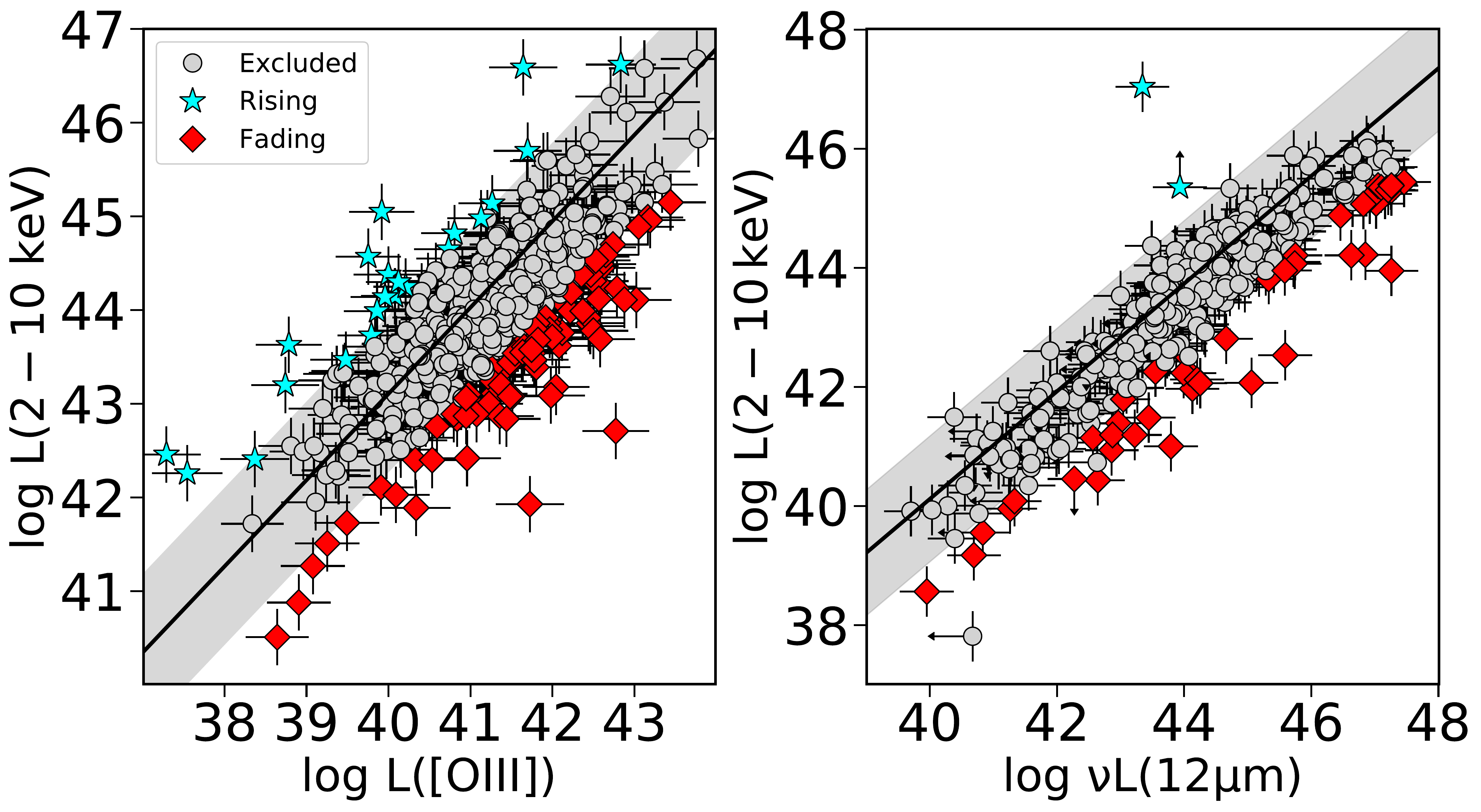}
    \caption{(left): The 2-10 keV X-ray luminosity, $\rm{L(2-10\,keV)}$, versus the [OIII] reddened corrected luminosity, $\rm{L([OIII])}$ (both in logarithmic scale), for the BASS sample reported by \citet{Berney15}. (right): The 2-10 keV X-ray luminosity, $\rm{L(2-10KeV)}$, versus the MIR 12$\rm{\mu m}$ continuum luminosity, $\rm{\nu L(12 \mu m))}$ in logarithmic scales. The black continuous line shows the best linear fit using type-1 AGN in the left panel, whereas it shows the relation reported by \citet{Gandhi-09} in the right panel. The gray shadowed region indicates the 2-$\rm{\sigma}$ from the linear relation. Gray circles, red diamonds, and cyan stars show the excluded, AGN rising candidates, and AGN fading candidates, respectively.}
    \label{fig:scalingrelation}
\end{figure*}

\section{Scaling relations and initial sample}\label{sec:scalingrelations}

We look for rising/fading AGN using well known linear relations between signatures of several AGN components at different wavelengths. This assumes that a relation is found when the involved components are traced by the same bolometric luminosity. Objects showing values out of these relations imply different bolometric luminosities for each component, and therefore, being candidates to a long term evolution of the bolometric luminosity for the source.

\subsection{The X-ray versus [OIII] \texorpdfstring{$\lambda$}{lambda-}5007 luminosity relationship}

We considered all sources classified as Seyferts reported by \citet{Berney15} to explore the relationship between X-ray and [OIII] luminosities. This relation is of the form:
\begin{equation}
    \rm{log \ L(2-10keV) = \alpha log L([OIII]) + \beta}
\label{eq:lx2loiii}
\end{equation}

This relationship is well explored, showing a good behaviour from high- to low-luminous AGN \citep[e.g.][]{Ward88,Panessa06,Gonzalez-Martin09B}. The X-ray luminosity traces the accretion disk associated with the current bolometric luminosity. The $\rm{L([OIII])}$ traces the NLR with a bolometric luminosity associated with this kpc-scale structure. Note also that the [OIII] emission parent ion recombines much more rapidly than almost any other \citep[especially hydrogen recombination, whose timescale can be thousands of years in Extended Emission-Line Regions - EELR, ][]{Binette1987}. Meanwhile, the X-ray emission traces the nuclear source luminosity associated with $\rm{\sim}10^{-2}$ pc scale structure or timescale of $\rm{\sim} 3\times 10^{-2}$ yr \citep{Hawkins07}.

We used the optical and X-ray measurements reported by \citet{Berney15}. Their optical data are taken from the BAT AGN Spectroscopic Survey (BASS) Data Release 1 \citep[][]{Koss17}. The BASS catalog contains 67.6$\%$ of the total AGN detected in the \emph{Swift} BAT 70-month catalog and has an average redshift of $z = 0.10$. The optical measurements were obtained using a combination of power law plus Gaussian components to fit the continuum and the emission lines, respectively. The flux uncertainty for the [O~III] emission line is typically below 0.01 \%. We use the intrinsic fluxes reported by them which were corrected for host galaxy extinction using the Balmer decrement (i.e. $\rm{H\alpha/H\beta}$). They corrected for extinction using the narrow Balmer line ratio $\rm{H\alpha/H\beta}$ assuming an intrinsic ratio of R = 3.1 \citep[e.g.][]{Ferland86} and the reddening curve provided by \citet{Cardelli89}. The 2-10~keV intrinsic fluxes are based on a homogeneous spectral fitting using the best available X-ray data with simultaneous fitting of the 0.2-10~keV band (from \emph{XMM-Newton}, \emph{Chandra}, or \emph{Swift}/XRT) and the 14-195~keV band from \emph{Swift BAT} \citep[details in][]{Ricci17}. Our sample contains 579 sources with [OIII] and X-ray measurements; one type BL LAC, 55 type-1, 107 type-1.2, 100 type-1.5, 96 type-1.9, and 220 type-2 AGN.

In order to minimize issues due to attenuation and/or obscuration of type-2 AGN, we define the slope and offset of this relationship (Eq. \ref{eq:lx2loiii}) using unobscured type-1 AGN only. The resulting relationship together with the data are shown in Fig.\,\ref{fig:scalingrelation} (left panel). We obtained a slope of $\rm{\alpha= 0.92}$ and an offset of $\rm{\beta = 6.27}$, which are consistent with previous results. We selected rising/fading AGN as the sources that are located outside 2$\sigma$ from this relationship (shaded area in Fig\,\ref{fig:scalingrelation}). In total, we obtained 113 candidates using this criterion; 23 rising and 90 fading sources. Among them we found 69 type-1, 43 type-2, and one BL LAC\footnote{This source could have errors in the [OIII] line flux measurement. It will be discarded in the next section through other criteria.}.

\vspace{0.5cm}
\subsection{MIR versus X-ray luminosity relationship}

The X-ray versus MIR luminosity relationship is also a very well known scaling relation in AGN \citep[][]{Elvis1978, Glass1982, Krabbe2001, Lutz2004, Ramos2007} of the form:
\begin{equation}
    \rm{ log \nu L(12\mu m) = \alpha log \ L(2-10keV) + \beta} \label{eq:lx2lmir}
\end{equation}

It has been interpreted as a connection between the accretion disk and the dusty torus. \citet{Gandhi-09} reported $\rm{\alpha = 1.11\pm0.07}$ and $\rm{\beta = -4.37\pm3.08}$ using a sample of 42 AGN, with a median $z = 0.1$ and a range of MIR luminosities of $\rm{\log (\nu L_{12 \mu m}) =[41.4-44.6]}$. Posterior analysis have shown consistent values for these constants \citep[e.g.][]{Asmus-15}. At the typical range scale of the dusty torus of $\rm{\sim 0.3 - 10}$\,pc, this structure traces the bolometric luminosity of the source roughly from 10 - 30\,yr \citep{Lyu19}. Thus, outliers in this relation might trace changes in shorter times scales than the X-ray versus [OIII] relation explained above.

We use all the sources reported in \citet{Asmus-15}, which contains a catalog of 253 sources with ground-based MIR photometric data from several observatories (e.g. VIRIS/VLT, T-ReCS/Gemini, CanariCam/GTC, Michelle/Gemini). This sample contains AGN with MIR luminosities of $\rm{log (\nu L_{12\mu m} ) =  [39.7 -45.7] }$ and redshifts lower than $\rm{z<0.4}$. \citet{Asmus-14} mark sources as non-reliable for low count-rate observations, or AGN classified as Compton-thick obscured sources, for which X-ray observations with \textit{XMM}-Newton, \textit{Suzaku}, or \textit{NuSTAR} were not available at the time of that publication. We do not exclude these sources to avoid losing potential candidates for fading/rising AGN activity. We further investigate the reliability of these candidates in Section\,\ref{sec:candidates}.

We also added to the analysis the sources reported by \citet{Stern-15}, which includes several samples in order to compare with high-luminosity AGN (magnitude in the I band in the range $\rm{I~[-29.3, -30.2]}$) and high redshift AGN ($1.5 < z < 4.6$). Among these samples they include the mixed \textit{Fifth Data Release Sloan Digital Sky Survey/XMM-Newton Quasar Survey} \citep[SSDS DR5][]{Young-09}, the \textit{Serendipitous Extragalactic X-ray Source Identification} \citep[SEXSI][]{Harrison-03, Eckart-05, Eckart-06, Eckart-10} sample, the QSO sample from \citet{just07}, and both Compton-thick and Compton-thin samples using \textit{NuSTAR} data, with MIR luminosities from \emph{WISE}, VLT/VISIR and \emph{Spitzer}.

Note that the intrinsic (i.e. absorption corrected) 2-10 keV X-ray luminosities included in these samples are obtained from literature. We refer the reader to \citet{Asmus-15} and \citet{Stern-15} for further details. We further explore in Section\,\ref{sec:robustness} if the line-of-sight absorption correction is robust, with particular attention to the Compton-thick nature of the sources to provide a more robust list of rising/fading AGN candidates.

Altogether, the sample we explore includes 419 AGN; 224 type-1 AGN, 123 type-2 AGN, 32 LINERS, 37 composite AGN, and three unclassified AGN. Among them, 253 sources come from the sample reported by \citet{Asmus-14} and 166 sources from the sample published by \citet{Stern-15}). Fig.\,\ref{fig:scalingrelation} (right panel) shows the MIR versus X-ray luminosities for the combined sample and the relation found by \citet{Gandhi-09}. As in the previous section, we consider as rising/fading candidates AGN those sources outside of this relation at the 2-$\sigma$ level. Note that, although two objects fall out of the relation, they have been excluded from the sample because the MIR luminosity is an upper-limit and therefore, they are consistent with being in the relation. In total, we obtained 49 sources: Two rising and 47 fading. Among them we found 17 type-1 and 18 type-2 AGN, 10 LINERs, and four composite AGN.

\begin{figure*}
\begin{center}
\includegraphics[width=0.69\columnwidth]{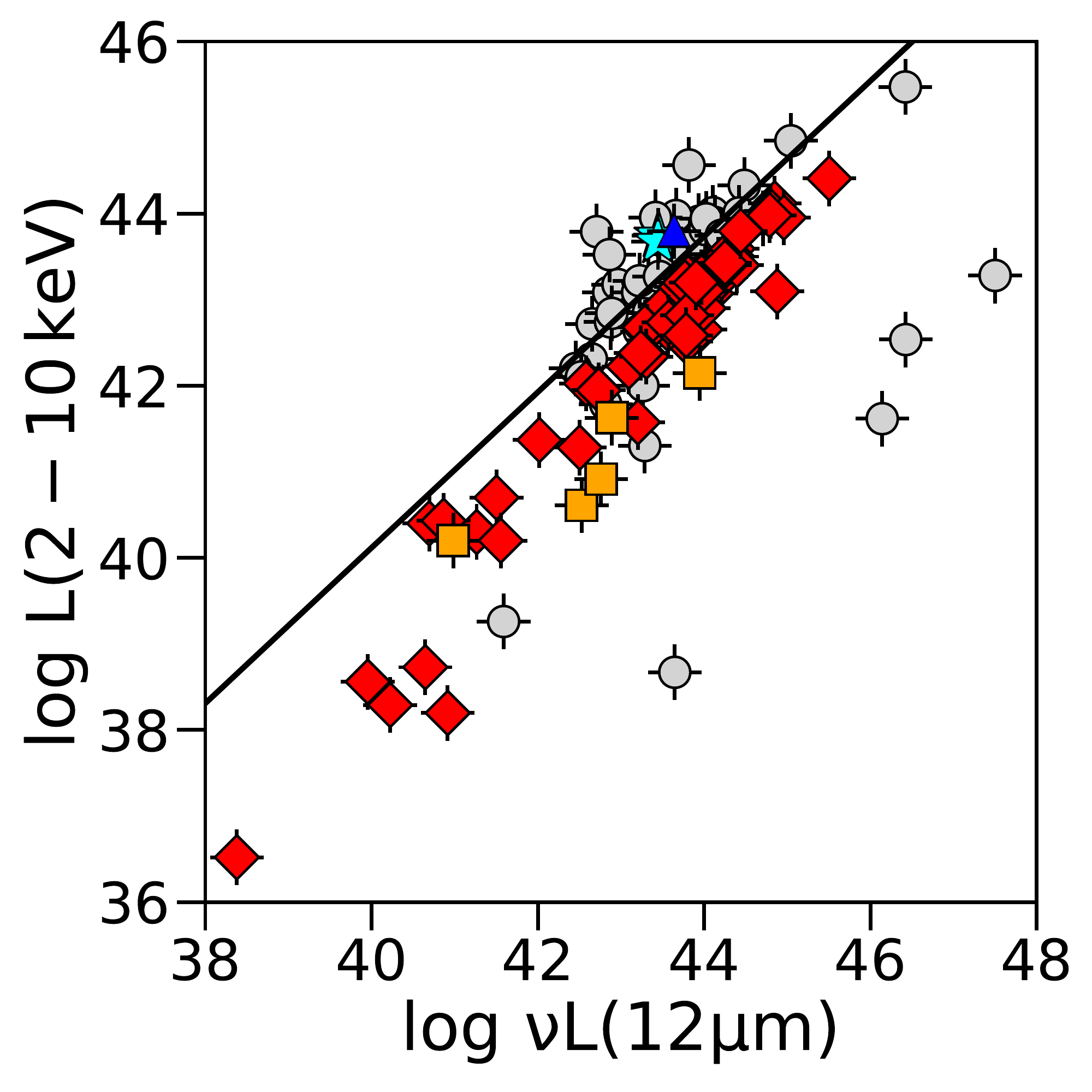}
\includegraphics[width=0.69\columnwidth]{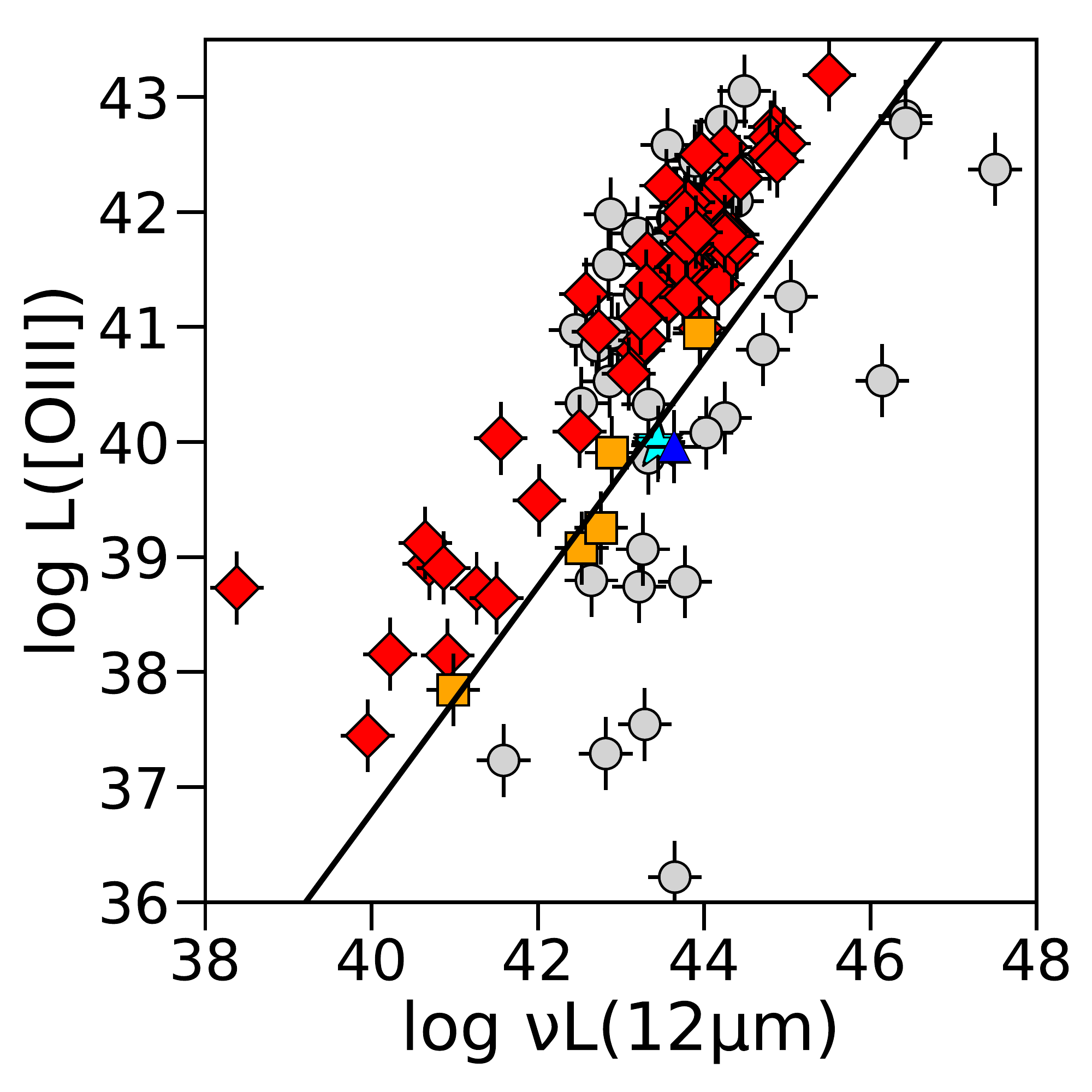}
\includegraphics[width=0.69\columnwidth]{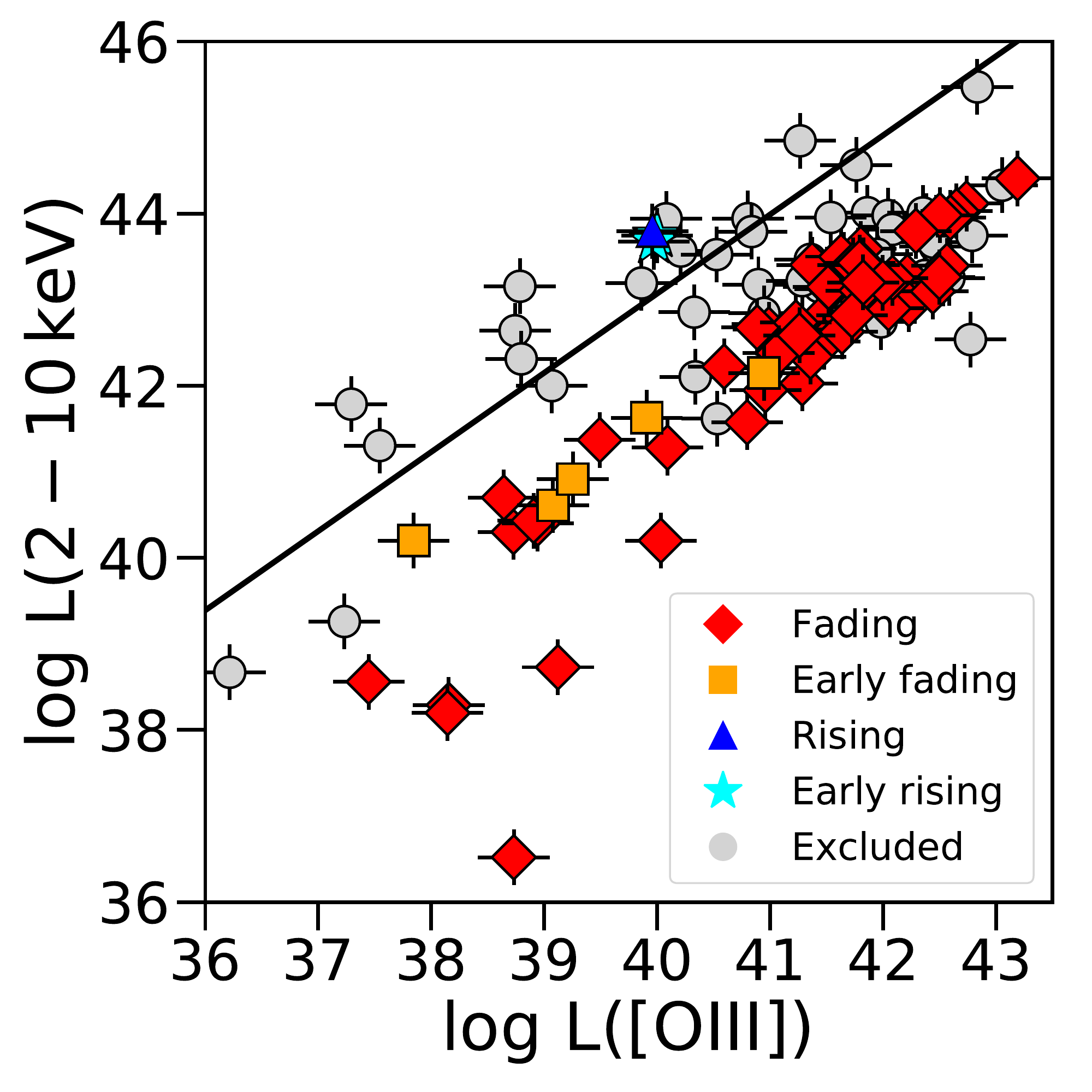}
\includegraphics[width=0.69\columnwidth]{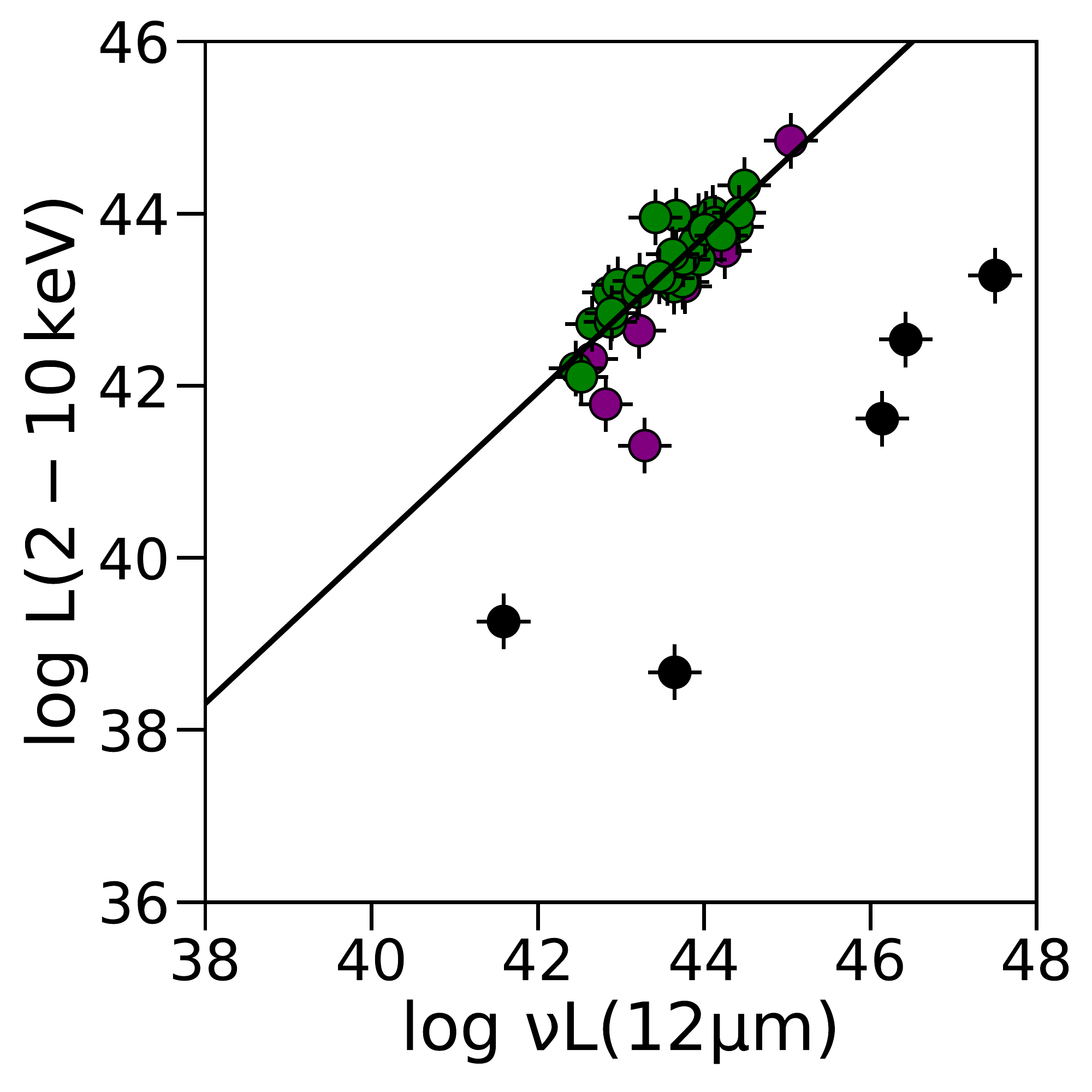}
\includegraphics[width=0.69\columnwidth]{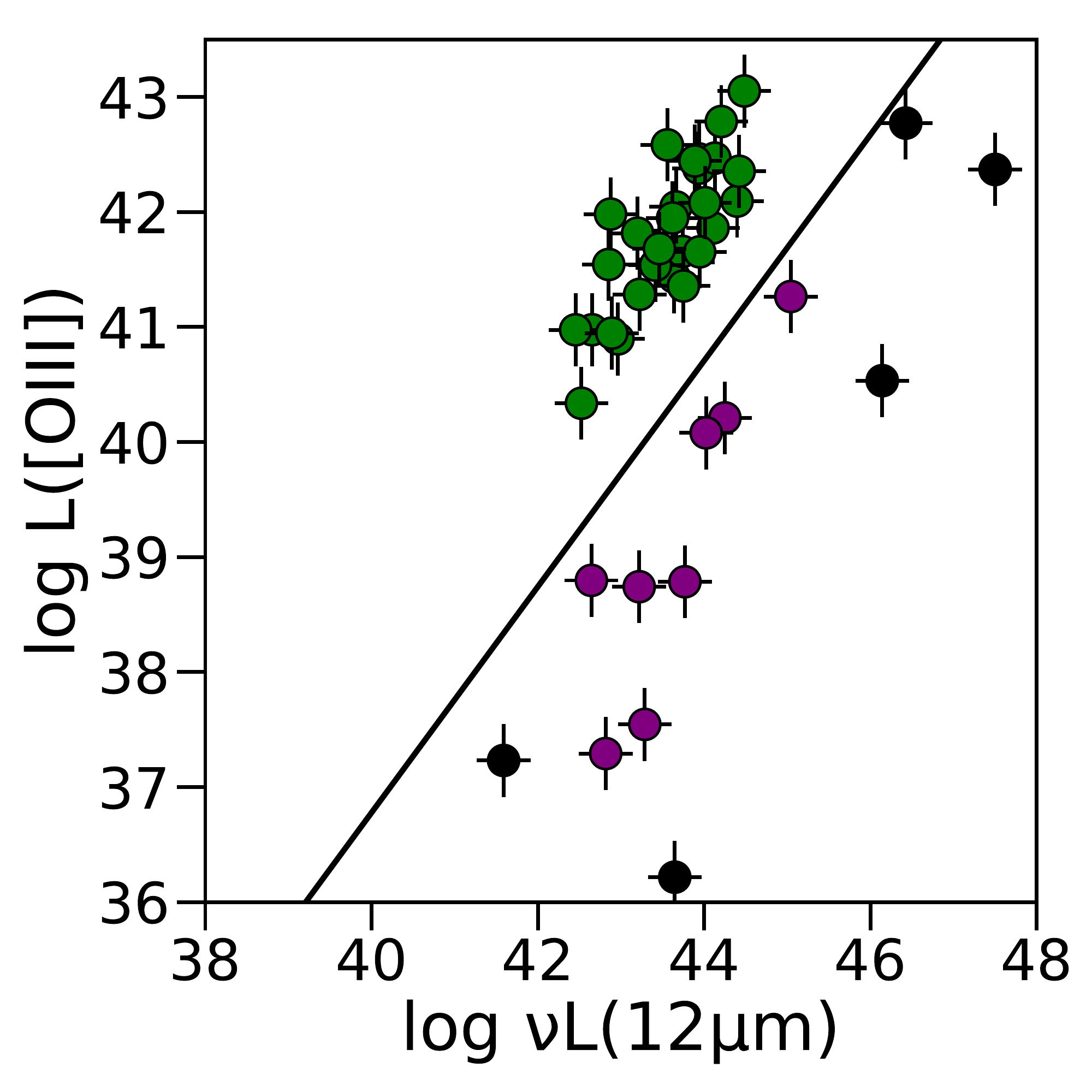}
\includegraphics[width=0.69\columnwidth]{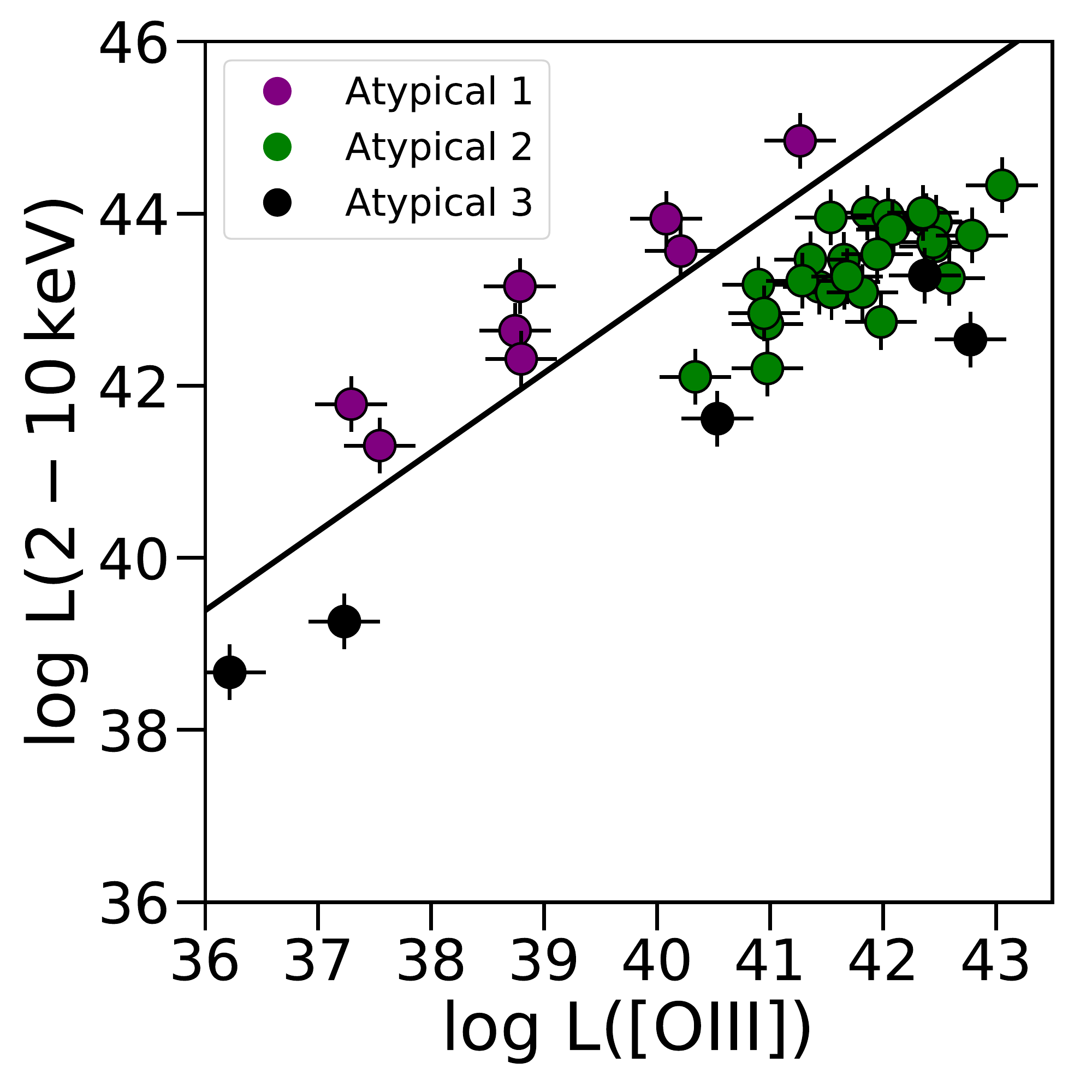}
\caption{Scaling relations used to confirm (top panels)  and reject (bottom panels) the fading and rising candidates: X-ray luminosity versus MIR luminosity (left panels), [OIII] luminosity versus MIR luminosity (middle panels), and X-ray luminosity versus [OIII] luminosity (right panels). The continuous black lines in the left and right panels are those used to select objects in Section\,\ref{sec:scalingrelations}. The continuous black line shown in the middle panel is obtained by combining the MIR to X-ray and the [OIII] to X-ray scaling relationships. Gray dots are rejected candidates in the top panels, red diamonds are 'Fading' candidates, orange squares are 'Early fading' candidates, blue triangles are 'Rising' candidates, and cyan starts are 'Early rising' candidates in the top panels. Atypical classification is shown in the bottom panel as purple, black, and green dots for the 'Atypical 1', 'Atypical 2', and 'Atypical 3', respectively (see text).}
\label{fig:candidates}
\end{center}
\end{figure*}

\section{AGN fading/rising candidate sample}\label{sec:candidates}

In summary, we select a total of 137 fading AGN and 25 rising AGN candidates using the two selection criteria explained in Section \ref{sec:scalingrelations}. Among them, one object was in common in both selection criteria. Therefore, our initial sample contains 161 candidates.

However, using only a single criterion is not enough to consider the candidate as secure. AGN are variable sources by definition. These variations are expected to occur throughout the entire electromagnetic spectrum. Inner components as the accretion disk are expected to vary in time scales of hours. Thus, random variations of the disk are expected to occur when compared to the torus or the NLR, without implying a consistent fading/rising scenario. These variations will reflect into the scaling relations above as scatter. Some of this scatter could be included in our sample of fading/rising candidate. Thus, the three AGN components should show consistent fading/rising behaviour of the bolometric luminosity for the AGN to be considered as a good candidate. For this reason we complete our compilation of X-ray, MIR, and [OIII] luminosities from literature. We only use $\rm{L([OIII])}$ corrected from reddening (mainly from SDSS) and the MIR luminosities are obtained from several catalogs at the 12$\rm{\mu m}$, using as main search engine the Nasa Extragalactic Database (NED\footnote{http://ned.ipac.caltech.edu.}) \citep[e.g.][]{Risaliti99,Tran03,Heckman05, Goulding09,Lamastra09, Noguchi10, Jin12, Berney15}. We complete the three luminosities for 110 objects. Among them, 58 are type-1 AGN and 52 are type-2 AGN.

Although we already had X-ray luminosities for all the candidates, we took particular care to look for X-ray luminosities fully corrected from obscuration along the line of sight. Indeed, uncorrected luminosities in moderate to highly obscured AGN ($\rm{N_H > 10^{23}cm^{-2}}$), could wrongly locate the object outside the scaling relations. This might overestimate the detection of fading AGN candidates. All the X-ray measurements used in previous section are intrinsic luminosities (i.e. corrected from obscuration). However, many of them rely on spectral analysis below 10 keV. This might wrongly estimate the intrinsic luminosity for Compton-thick AGN (with $\rm{N_H > 3\times 10^{24}cm^{-2}}$). X-ray spectra with energies below 10 keV cannot be used to estimate the true value for the obscuration if that is above the Compton-thick limit \citep{Comastri04}. In order to mitigate this effect, we look for obscuration measurements for all the objects, giving priority to the analysis where spectra above 10 keV is considered (e.g. \emph{NuSTAR} or \emph{Suzaku}). Most of them were obtained from \emph{Swift} BAT 70-month catalog \citep[][]{Ricci17}. We also check for signatures of Compton-thickness reported in the literature and found the $\rm{N_{H}}$ measurements for all but 35 objects. We also found archival \emph{NuSTAR} data for 18 out of these 35 objects. The \emph{NuSTAR} spectra were extracted using standard procedures and fitted to power-law model with partial covering to estimate the $\rm{N_{H}}$ and X-ray intrinsic luminosity (see Appendix\,\ref{sec:appendixA} for more details). 

\begin{figure}
\begin{center}
\includegraphics[width=1.0\columnwidth]{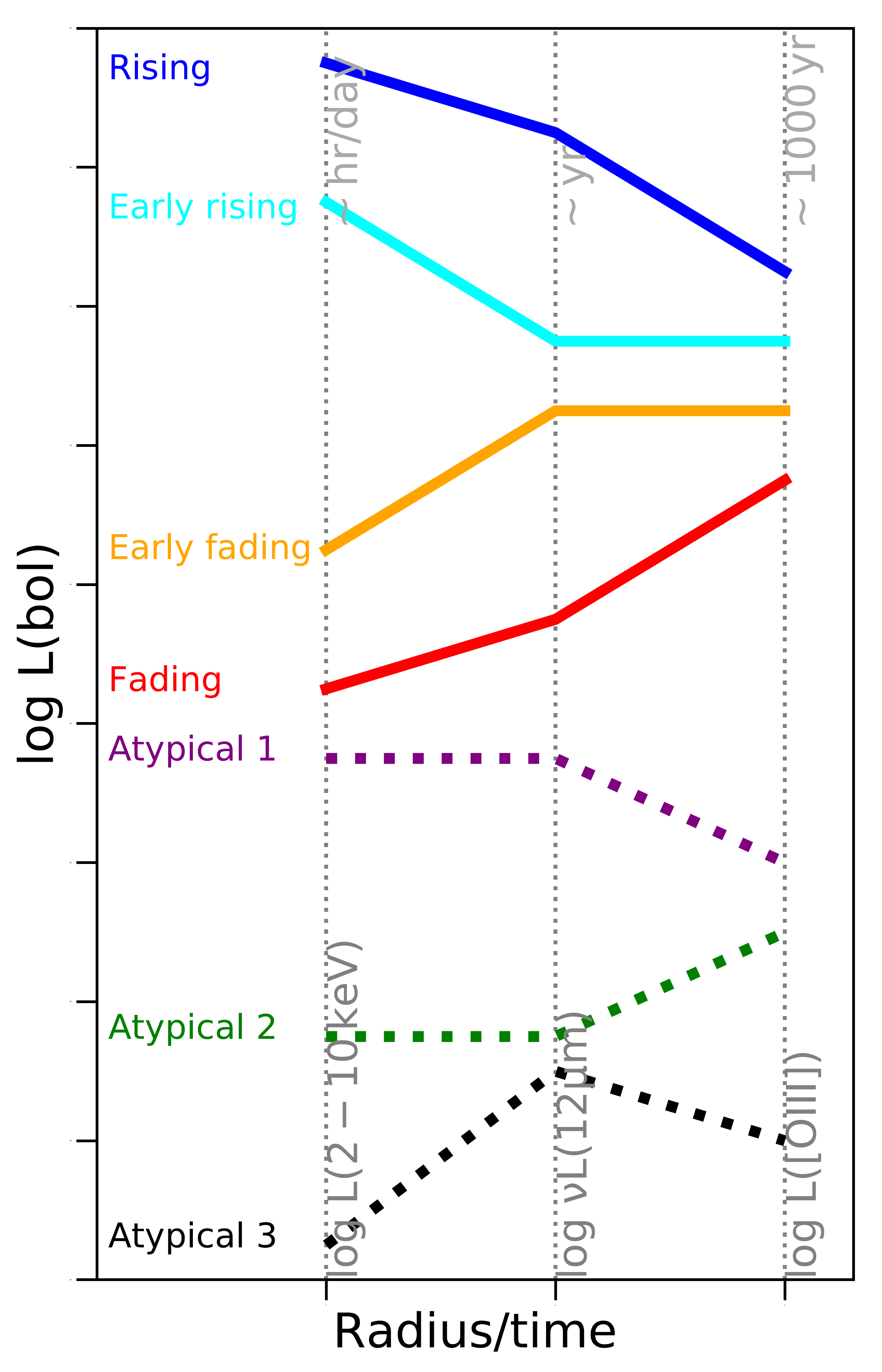}
\caption{Illustrations of the different types of classified objects according to the bolometric luminosity obtained with X-ray, MIR, and [OIII] luminosities (in logarithmic representation). These bolometric luminosities correspond to hr/days, yr, and 1000 of yr (also highlighted in the top horizontal axis of the plot). Fading and rising candidates are shown with continuous lines. The behaviours presented for the atypical objects are shown with dashed lines (see text for more details). Axes in arbitrary units.}
\label{fig:prototypes}
\end{center}
\end{figure}

Fig.\,\ref{fig:candidates} shows the three scaling relations for our 110 fading/rising candidates: X-ray versus MIR luminosity (left), [OIII] versus MIR luminosity (center), and X-ray versus [OIII] luminosity (right). The scaling relations used in Section\,\ref{sec:scalingrelations} are shown as a black continuous line in the left and right panels. Middle panel shows the expected relation combining the previous two relations. According to the position in this plot we classify most of the sources into four main categories (see top panels in Fig.\,\ref{fig:candidates}):

\begin{itemize}
\item \underline {Fading candidates} (red diamonds): the three diagrams show an increase of the bolometric luminosity from the disk toward the torus and NLR (red line in Fig.\,\ref{fig:prototypes} shows the estimated behavior of the bolometric luminosity for these sources). We found 53 objects belonging to this category.

\item \underline {Early fading candidates} (orange squares): the object shows an increase of the bolometric luminosity from the disk toward the torus and from the disk toward the NLR. However, the bolometric luminosity obtained for the torus and NLR are consistent to each other (orange line in Fig.\,\ref{fig:prototypes}). Five objects belong to this category.

\item \underline {Rising candidates} (blue triangle): the three diagrams show the object in a consistent decrease of bolometric luminosity from the disk toward the torus and the NLR (blue line in Fig.\,\ref{fig:prototypes}). Only one object belongs to this category.
\item \underline {Early rising candidates} (cyan stars): the object shows a decrease of the bolometric luminosity from the disk toward the torus and from the disk toward the NLR. However, the bolometric luminosity obtained for the torus and the NLR are consistent to each other (cyan line in Fig.\,\ref{fig:prototypes}). Two objects belong to this category.
\end{itemize}

We keep early fading/rising candidates into the sample under the interpretation that these objects might be a premature fading/rising of the central source, still not clearly shown in the outskirts of the system. Note that the object is considered above/below the relation with $\rm{\Delta log(L)>0.2}$, which is consistent with the systematic errors in these relations. Table\,\ref{table:candidates} compiles the names and general information for these objects. 

Among the 110 objects studied, 49 have been rejected on the basis of non expected behaviour on the set of X-ray, MIR, and [OIII] luminosities. Eight of these objects show a monotonic increase or decrease on the luminosities but this behaviour is inconclusive due to $\rm{\Delta log(L)<0.2}$. Therefore, these eight sources are rejected. We classify the 41 remaining sources into three subcategories (see bottom panels in Fig.\,\ref{fig:candidates}):
\begin{itemize}
    \item[-] \underline{Atypical 1} (purple points): eight objects show a decrease on the bolometric luminosity between the disk and the torus and between the disk and the NLR. However, the torus shows similar bolometric luminosity compared to that of the disk, which might be inconsistent with the rising scenario (shown as dotted-purple line in Fig.\,\ref{fig:prototypes}).
    \item[-] \underline{Atypical 2} (green points): 28 objects show an increase of the bolometric luminosity associated with the NLR when compared to that of disk/torus that might indicate a fading of the source. However, there might not be a consistent fading scenario when it comes to the comparison between torus and disk bolometric luminosity (shown as dotted-green line in Fig.\,\ref{fig:prototypes}).
    \item[-] \underline{Atypical 3} (black points): five objects show an increase of the bolometric luminosity for the disk when compared to that of the torus and also for the disk compared to that of the NLR. However, there is a decrease on the bolometric luminosity of the NLR compared to that of the torus, which might be inconsistent with the fading scenario (shown as dotted-black line in Fig.\,\ref{fig:prototypes}).
\end{itemize}

\noindent These 41 candidates showing atypical behavior are included in Tab.\,\ref{table:rejected}.

\begin{figure}[!th]
\begin{center}
\includegraphics[width=1.\columnwidth]{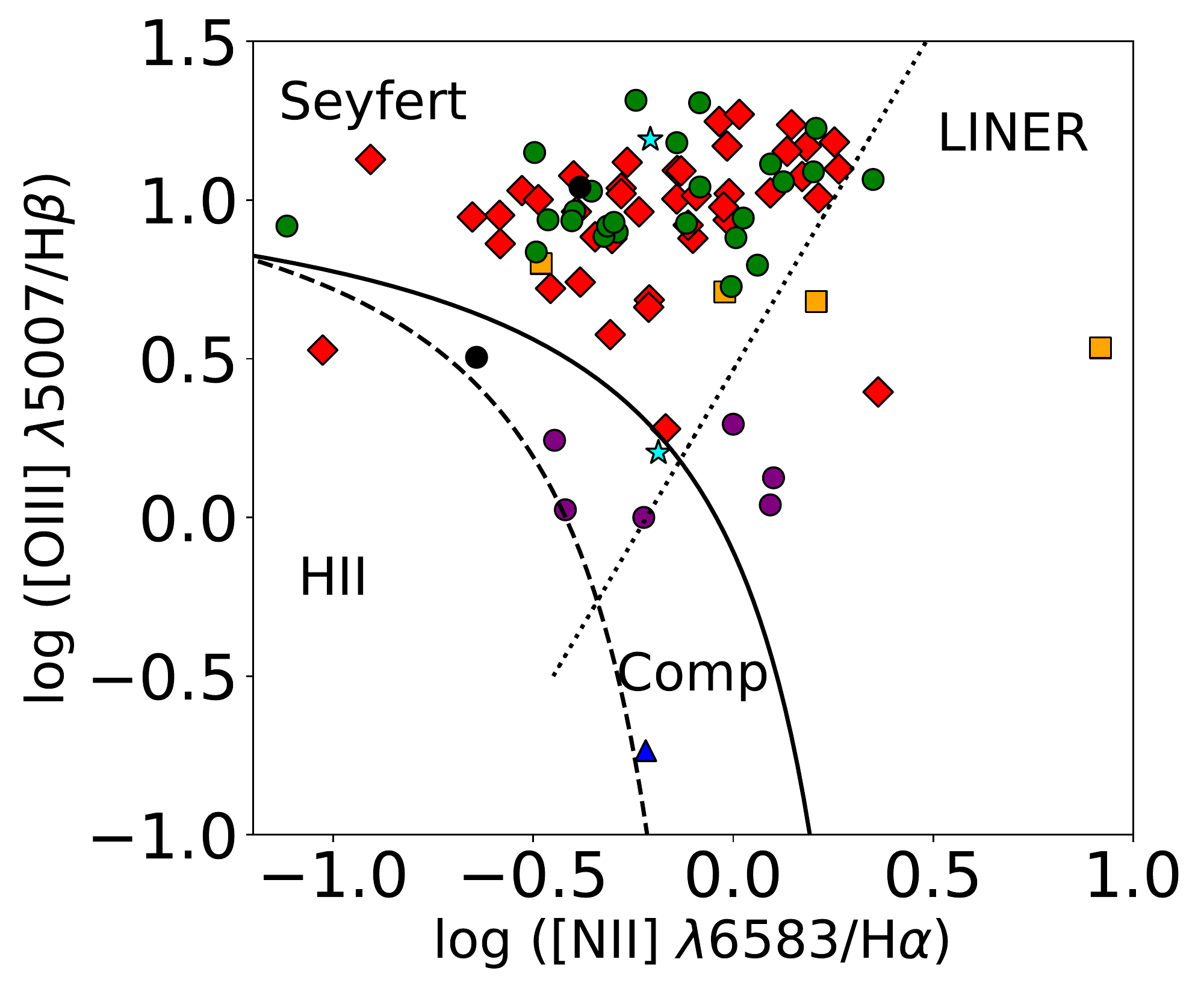}
\caption{[O\,III]/H$\rm{\beta}$ versus [N\,II]/H$\rm{\alpha}$ emission-line ratios. The solid line show the upper limit for star-forming galaxies (or HII galaxies) proposed by \citet{kewley01}. The dashed line is the lower limit for AGN according to \citet{kauffmann2003}. The dotted line marks the separation between pure LINERs from Seyferts from \citet{kauffmann2003}. The symbols and color code is the same as that explained in Fig.\,\ref{fig:candidates}.}
\label{fig:BPT}
\end{center}
\end{figure}

\section{Robustness of the fading/rising candidates}
\label{sec:robustness}

We further explore here fading, rising, and atypical candidates (102 objects) to investigate the robustness of the methodology using ancillary observations available for the sample.

\subsection{AGN nature and [OIII] as tracer of the NLR}

The absence of the accretion disk could be considered as an indication of the switch-off of the nucleus if the torus and/or NLR are still present. However, it could also indicate that the object does not harbor an AGN, which is a particularly relevant discussion for low-luminosity objects with $\rm{L_{bol}< 10^{42}erg/s}$. Indeed, 23 objects belong to this category among our candidates.

In order to study the AGN nature of these sources, Figure\,\ref{fig:BPT} shows the $\rm{[O\,III]/H\beta}$ versus $\rm{[N\,II]/H\alpha}$ emission line ratios for 84 sources of our sample. This is a well known AGN diagnostic diagram firstly explored by \citet{Baldwin81} (details on the construction of this diagram are included in Appendix\,\ref{app:bpt_diagram}). We only found one object (Mrk\,335, a fading candidate) not consistent with pure AGN according to the limit proposed by \citet{kauffmann2003}. However, this is a well-known and bright AGN from the X-ray point of view \citep[e.g.][]{Parker19}. Furthermore, other six objects are not consistent with pure AGN according to the demarcation proposed by \citet{kewley01}. Among them ESO137$-$G034 belongs to atypical 3, 2MASSX\,J02420381$+$0510061, Cen\,A, and NGC\,3079 to atypical 1, 2MASSX\,J14391186$+$1415215 to early rising, and 2MASSX\,J08551746$-$2854218 to the rising class. This group is consistent with a composite behaviour of the source at optical wavelengths. Interestingly, two out of the three rising candidates are in this latter category. Note that the [OIII] line emission could be contaminated by star-forming processes for these seven objects. This would move the objects towards the left in Fig.\,\ref{fig:candidates} (right) and downwards in Fig.\,\ref{fig:candidates} (center). However, note that even if it were the case, both early rising and rising objects would still be classified as such, as they would still remain out of the expected correlations in Fig.\,\ref{fig:candidates}. Some of the sources could be affected by lower metallicity than the ``standard'' ones used to set up the various versions of the BPT strong-line diagram, where AGN-ionized gas can masquerade as ionized by stars. \citet{Groves06} show that the respective BPT models evaluations may return different metallicity values. If we consider models with 0.25 solar metallicty, all of our sources could be classified as AGN. However, in order to be conservative, we keep the AGN classification as that found when using solar metallicity which is the most restrictive classification.

\begin{figure}
\begin{center}
\includegraphics[width=1.\columnwidth]{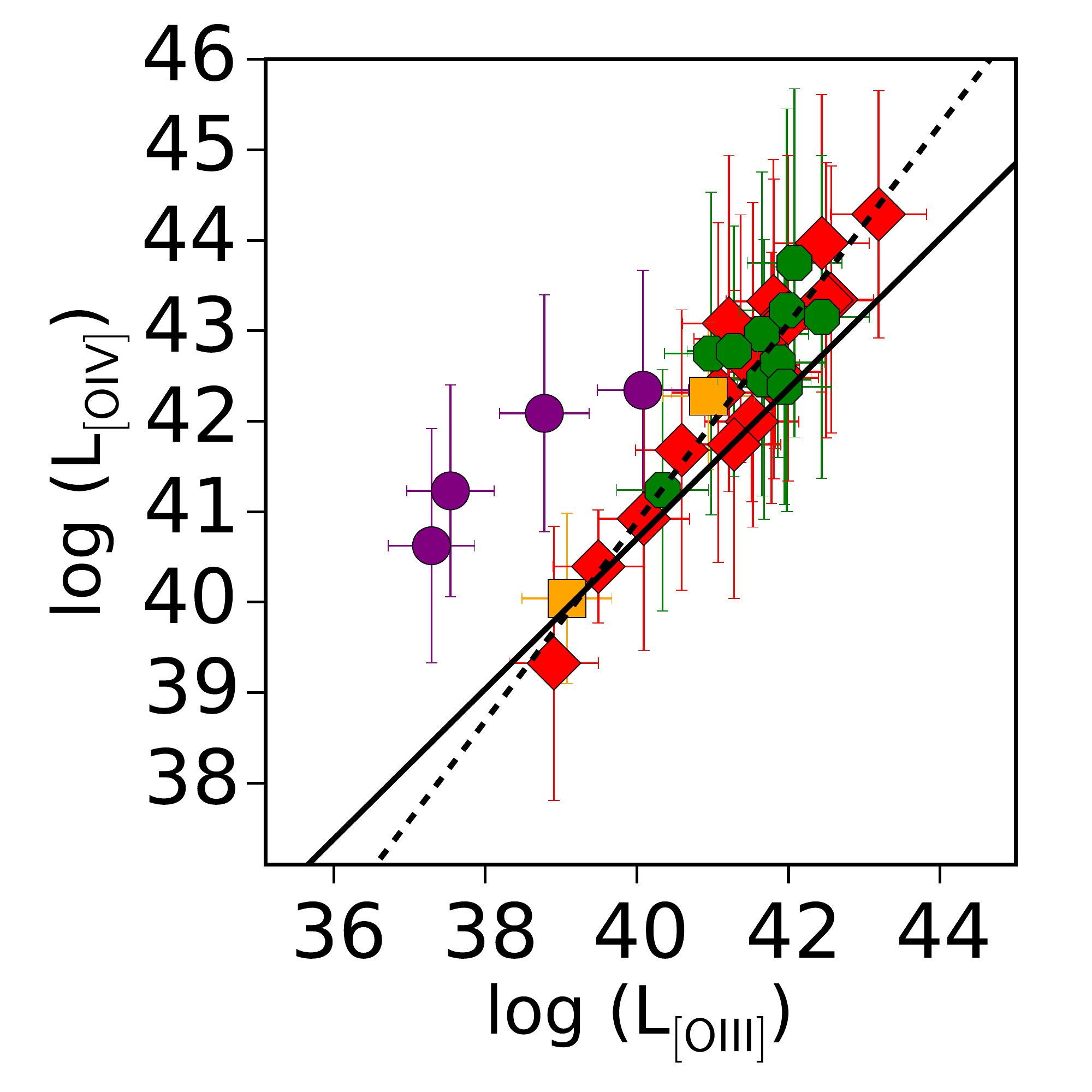}
\caption{[OIII] at $\rm{5007\,}$\AA\, luminosity, $\rm{L_{[OIII]}}$, versus [OIV] at $\rm{25.9\,\mu m}$ luminosity, $\rm{L_{[OIV]}}$. The black solid line represent the relationship found by \citet{Dickens14} $(\rm{log(L_{[OIV]}) = 0.83 \, log(L_{[OIII]})+7.5}$) and the dashed line represents the best fit to our data ($\rm{log(L_{[OIV]}) = 1.09 \, log(L_{[OIII]})-3.03}$). The symbols and color code is the same as that explained in Fig.\,\ref{fig:candidates}.}
\label{fig:LOIIIvsLOIV}
\end{center}
\end{figure}

Objects classified as atypical 1 (purple dots in Fig.\,\ref{fig:candidates}) tend to locate at lower $\rm{[O\,III]/H\beta}$ compared to other AGN in the sample. This is easily explained by large-scale extinction affecting the [OIII] line emission (see below and Fig.\,\ref{fig:LOIIIvsLOIV}). Another reason for the contamination of the [OIII] emission might come from tidal tails seen after merging processes which in addition can cause scatter in the [OIII] scaling laws. However, we find that the vast majority of the sources ($\sim92\%$) are located in the AGN area of the plot, supporting the AGN nature of them and the use of [OIII] as a tracer of the NLR.

We can also compare the [OIII] with other tracers of the NLR to look for wrong estimates of the NLR bolometric luminosity. Fig.\,\ref{fig:LOIIIvsLOIV} shows the $\rm{L([OIV])}$ versus $\rm{L([OIII])}$ relation. Details on [OIV] flux measurements are given in Appendix\,\ref{app:SpectralFit}. We show the relation found for broad-line radio galaxies (BLRG) by \citet{Dickens14} (black solid line). We also show the linear relation found using objects belonging to the fading category (black dashed line). Fading, early fading, and atypical 1 categories are consistent with the relation found for BLRG. Interestingly, the four atypical 1 objects (named NGC\,612, NGC\,3079, MCG\,+04-48-002, and Cen\,A) tend to show an excess of $\rm{L([OIV])}$ compared to the $\rm{L([OIII])}$. This indicates that the [OIII] emission might be suffering from extinction ([OIV] emission is much less likely to suffer dust extinction than the optical forbidden lines).  Indeed, all of them show column densities at X-rays with values $\rm{N_{H}>10^{23}cm^{-2}}$ and one of them is well within the Compton-thick regime. According to this, the $\rm{L([OIII])}$ might be a factor of 10 to 100 higher for the atypical 1 candidates (as Fig.\,\ref{fig:LOIIIvsLOIV} suggests). Therefore, they will move closer to the expected linear relations for AGN and we can safely remove atypical 1 from the raising/fading candidates.

\vspace{0.8cm}
\subsection{X-ray luminosity as a proxy of the accretion disk bolometric luminosity}

\begin{figure*}
\begin{center}
\includegraphics[width=1.0\columnwidth]{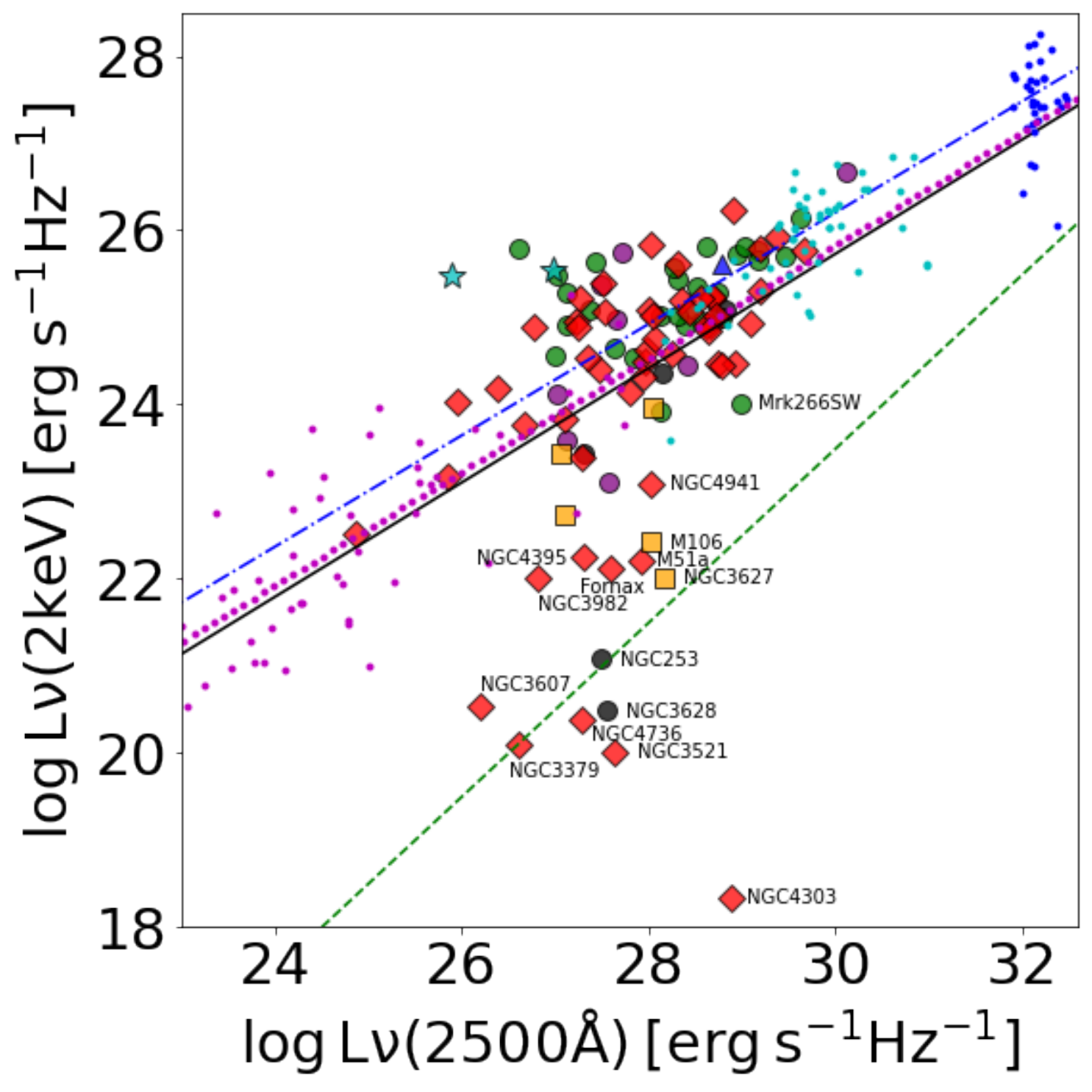}
\includegraphics[width=1.0\columnwidth]{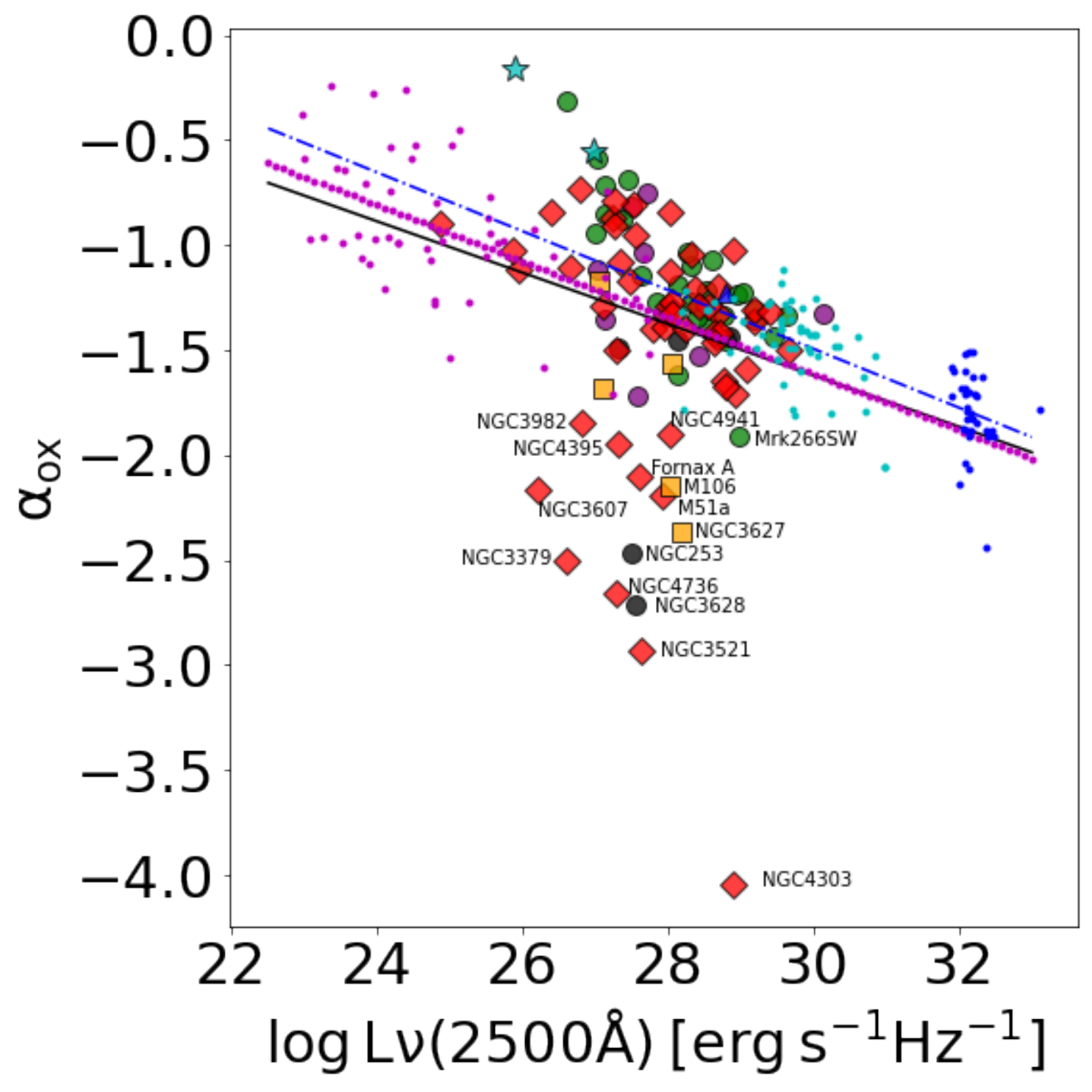}
\caption{(left): Relationship between log $\rm{L_\nu(2500}$\,\AA) vs log $\rm{L_\nu(2keV)}$ for fading/rising and atypical candidates. (right): Relationship between $\rm{\alpha_{ox}}$ vs log $\rm{L_\nu(2500}$\,\AA). The solid black line in both panels shows the best-fit for our sample. Blue, cyan, and magenta dots represent the high-luminosity \citep{just07}, intermediate-luminosity \citep{steffen} and low-luminosity AGN \citep{xu11} data. We compiled data from NASA/NED the optical/UV continuum luminosity or even we used extrapolated data if there are not data.}
\label{fig:XvsUV}
\end{center}
\end{figure*}

We explore here if the X-ray luminosity is a good tracer of the disk component by comparing it with optical/UV continuum emission where the peak of the disk luminosity occurs. For that purpose we compiled from NED the optical/UV continuum luminosity, $\rm{ log\, L_\nu(2500}$\,\AA), for the 102 candidates (see Col.\,11 in Tabs.\,\ref{table:candidates} and \ref{table:rejected}, respectively). Fig.\,\ref{fig:XvsUV} (left) shows $\rm{ log\, L_\nu(2500}$\,\AA) versus $\rm{log(L_\nu(2\,keV))}$. \citet{tananbaum79} define a relationship between the rest-frame monochromatic luminosity of the form $\rm{\alpha_{ox}=0.38~ log\,(L_{\nu}(2\,keV)/L_{\nu}(2500\, \textup{\r{A}}))}$, which links the accretion disk emission to that of the hot corona emission. The upper allowed limit for this ratio is $\rm{\alpha_{ox}=-2.5}$ (green dashed line in Fig.~\ref{fig:XvsUV}, left) for high luminosity AGN \citep{martocchia17}. We also include, for comparison purposes, as blue, cyan, and magenta dots the data and their best-fit  for high-luminosity \citep{just07}, intermediate-luminosity \citep{steffen}, and low-luminosity AGN \citep{xu11}, respectively.

Fig.\,\ref{fig:XvsUV} (right) also shows the distribution of $\rm{\alpha_{ox}}$ as a function of extinction-corrected 2500\,\AA\, monochromatic luminosity, $\rm{ log\, L_\nu(2500}$\,\AA). The black solid line is our best fit regression curve. In the left panel this linear relation is described by $\rm{\alpha_{ox} = -0.12}~\rm{log\,L_{\nu}(2500}$\,\AA)+ $\rm{2.04}$, consistent with Eq.~4 from \citet[]{xu11}. Similarly to the left panel, we also include the linear regression found by \citet{xu11}, \citet{just07}, and \citet{steffen}, respectively. Although a large number of objects are consistent with previously found relations, they show a wide range of $\rm{\alpha_{ox}}$ in a narrow luminosity range. 

We find an average $\rm{\alpha_{ox}=-1.37}$. The range of $\rm{ log\, L_{\nu}(2500}$\,\AA) and $\rm{log\,L_\nu(2\,keV)}$, partially overlap with the sample presented by \citet{xu11}. Furthermore, the linear fit to our data (black solid line in Fig.~\ref{fig:XvsUV}, left) is $\rm{log\,L_\nu(2\,keV)}$ =  $\rm{(0.65\pm0.15)}\rm{~ log\, L_{\nu}(2500}$\,\AA)~+~$\rm{(6.01\pm4.24)}$, which is quite similar to that found from \citet[]{xu11} (see their Eq.~2). In our sample, 96\% of the objects are consistent with the linear relation found. Interestingly, the two early rising candidates tend to locate above the linear relation (Fig.~\ref{fig:XvsUV}, left) and flat slopes (Fig.~\ref{fig:XvsUV}, right). These flat slopes might indicate a fundamental change of the accretion process, which might be associated with the rising scenario. However, it could also be interpreted as an over estimate of the accretion disk luminosity when using the X-ray emission. This would put back these two objects into the general correlations in Fig.\,\ref{fig:candidates}. Under this interpretation, these two objects might not be good rising candidates after all. This X-ray emission sometimes has a non-negligible contribution from reprocessed material that should not be taken into account for the intrinsic disk luminosity. This could explain the slight excess of X-ray luminosity compared to the UV luminosity. Unfortunately, a detailed analysis of the intrinsic and reprocessed X-ray emission is not possible for these two sources due to the lack of high energy X-ray spectra. Only objects with high column densities, reaching the Compton-thick regime, are expected to have a large contribution of reflection component. This is indeed the case of NGC\,1194, one of the two early rising candidates.  

\begin{figure*}
\begin{center}
\includegraphics[width=1.0\columnwidth]{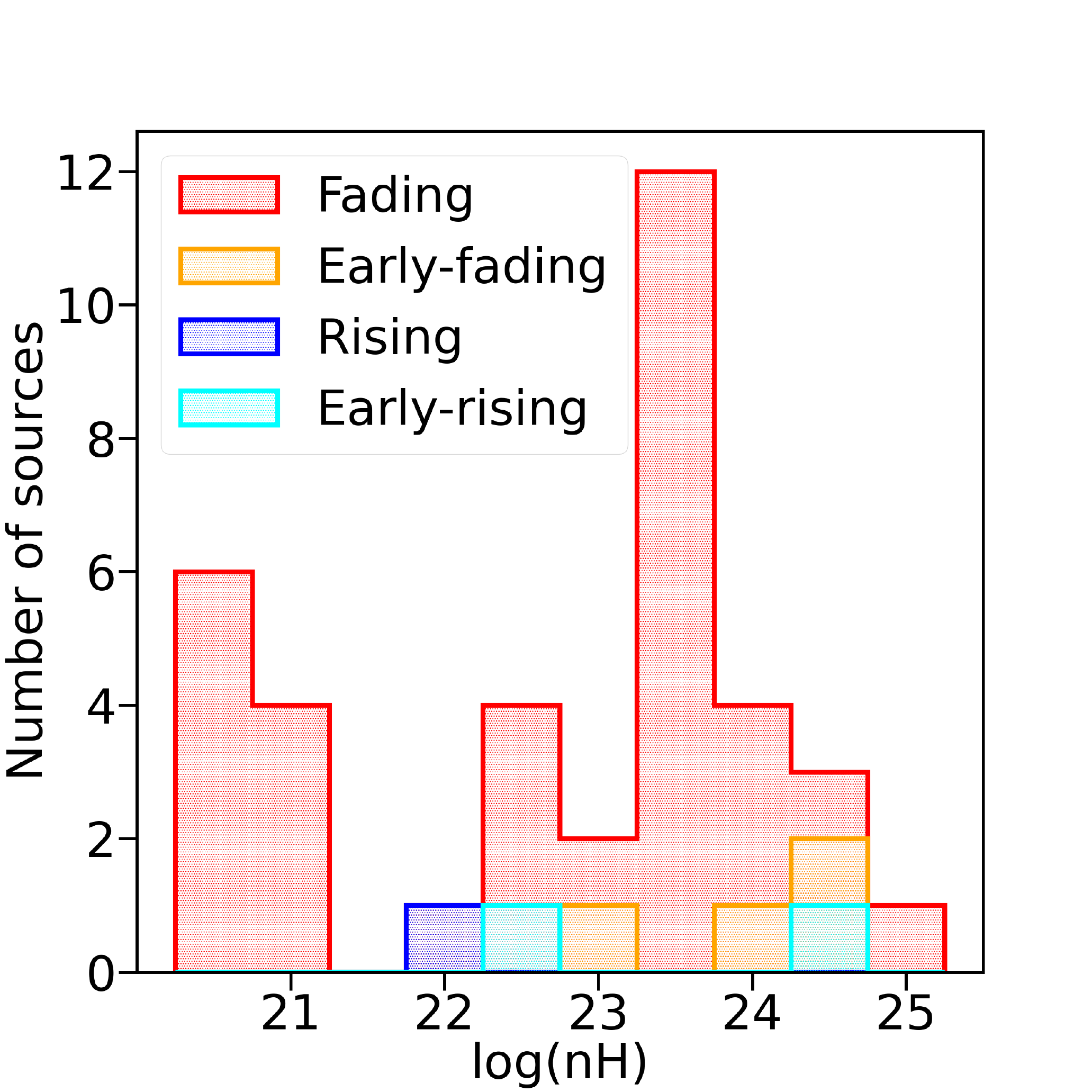}
\includegraphics[width=1.0\columnwidth]{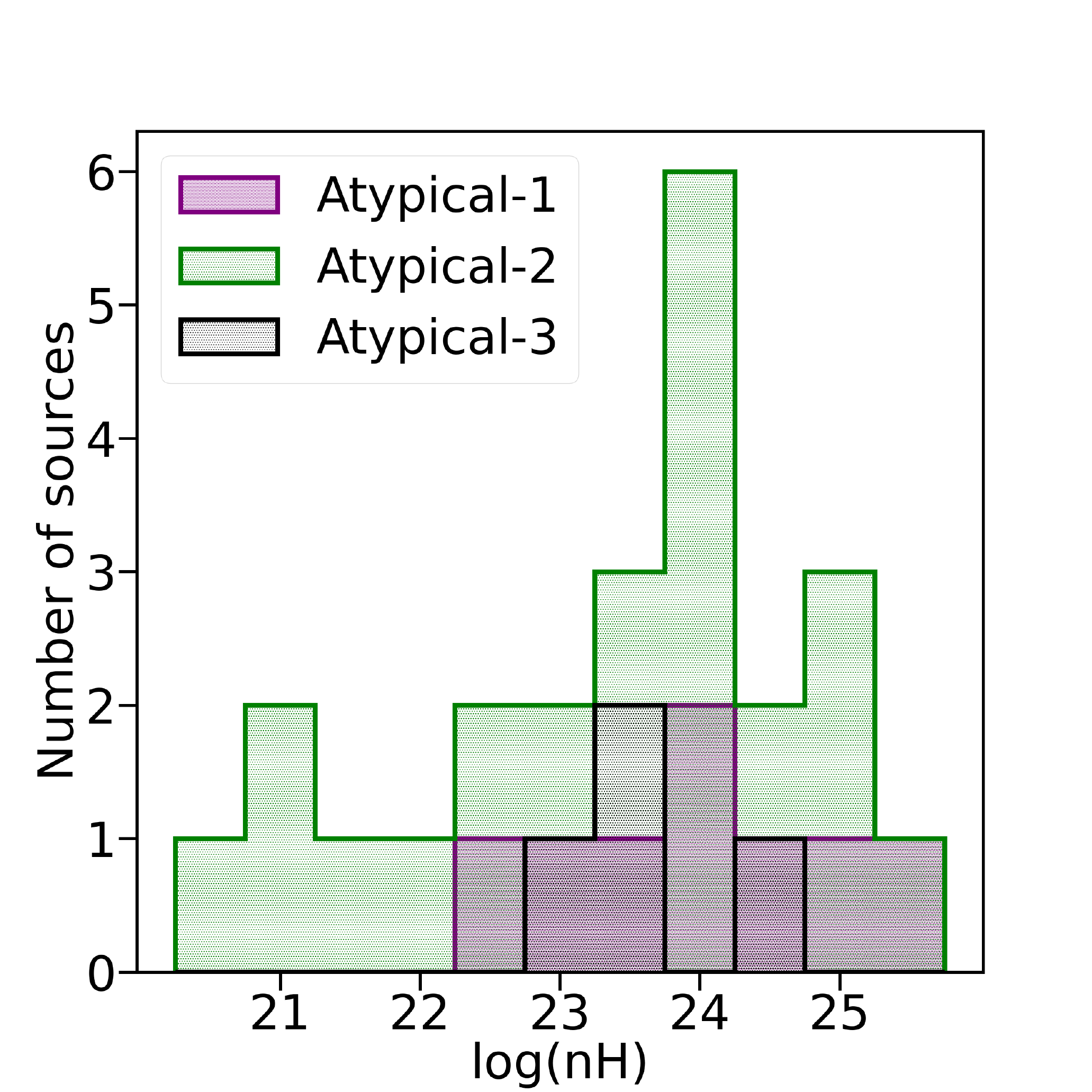}
\caption{Histograms of hydrogen column densities, $\rm{N_{H}}$ observed at X-rays for fading/rising AGN (left) and atypical AGN (right). It includes 37 fading, 4 early fading, 2 early rising, 1 rising, 23 atypical-2, 4 atypical-3, and 8 atypical-1.}
\label{fig:histograma_accepted}
\end{center}
\end{figure*}

Four sources are significantly below the $\rm{log\,L_\nu(2500}$\,\AA) versus $\rm{log\,L_\nu(2\,keV)}$ relation, with values below $\rm{\alpha_{ox}=-2.5}$ (namely NGC\,3521, NGC\,4303, NGC\,4736 and NGC\,3628). \citet{Gonzalez-Martin09A} classified NGC\,3628 as non-AGN at X-rays based on \emph{Chandra} extended morphology, the lack of the iron $\rm{K\alpha}$ emission line, and no radio jet found. Therefore, the extreme $\rm{\alpha_{ox}\sim-2.5}$ found for NGC\,3628 could be due to the lack of AGN at the center. However, we analysed the \emph{NuSTAR} spectrum of NGC\,3628 (see Appendix\,\ref{sec:appendixA}) finding a spectrum consistent with a mildly obscured AGN. Note, however, that this is a rather simplistic analysis and a reflection component has not been taken into account. The inclusion of such component might lead to different results. Indeed, Osorio-Clavijo et. al.\,(in prep.), analyse a sample of AGN with a complex model, accounting for a reflection component, finding for this particular source a significant reflection fraction ($\sim 50 \%$), without significant obscuration. \citet{Gonzalez-Martin09A} also classified NGC\,4736 as AGN showing a compact X-ray, UV and optical morphology and a jet contributing to the radio emission. Indeed, we also found a jet-like structure in NGC\,3521 while we classified the other three as diffuse emission (see discussion and Appendix\,\ref{sec:RadioImages}). BPT diagram in Fig.\,\ref{fig:BPT} confirms the AGN nature of NGC\,3521, NGC\,4303, and NGC\,4736 (there are not available data for NGC\,3628). An alternative explanation for the extreme discrepancy between $\rm{ L_\nu(2500}$\,\AA) and $\rm{L_\nu(2keV)}$ in these four sources is that the X-ray luminosity is not well corrected from absorption along the line of sight. We compiled the column densities (and computed them in some particular cases, see Appendix\,\ref{sec:appendixA}) preferring those where X-ray spectra include hard X-ray photons above 10\,keV. The distribution of $\rm{N_{H}}$ is reported in Fig.\,\ref{fig:histograma_accepted}. The wrong estimate of the $\rm{N_{H}}$ might explain the locus in Fig.~\ref{fig:XvsUV} for NGC\,3521, NGC\,4303, and NGC\,4736 because no spectra above $\rm{10\,keV}$ is available for these three objects, and such energy range is necessary to discard that the sources are in the Compton-thick regime. Therefore, spectra in UV region and above $\rm{>10\,keV}$ either with \emph{Suzaku}, \emph{NuSTAR}, or future X-ray facilities are needed to confirm the fading nature of these three sources. It is certainly not the case for NGC\,3628 where we correct the spectrum from its absorption thanks to \emph{NuSTAR} (see Appendix \ref{sec:appendixA}). 

Although the other sources are within the standard range of $-2.5 < \rm{\alpha_{ox}} < -1.5$, it is evident the scatter; it is large for Fornax\,A, M\,51a, NGC\,3379, NGC\,3607, NGC\,3627, Mrk\,266SW, NGC\,253, NGC\,3982, M\,106, NGC\,4395, and NGC\,4941 being well below the X-ray versus UV linear relation. The BPT diagram confirms the AGN nature of M\,106, NGC\,4395, NGC\,4941, but there are not available data for the others. Among them, NGC\,253, M\,51a, MRK\,266SW, and NGC\,3982 are known AGN (see NED). Based on X-ray observations \citet{Gonzalez-Martin09A} classified NGC\,3379, NGC\,3607, and NGC\,3627 as non-AGN. Interestingly, these three sources are right on the limit of $\rm{\alpha_{ox}}= -2.5$ (green dashed line in Fig.\,\ref{fig:XvsUV}). Among these three objects, we find that NGC\,3627 is consistent with a Compton-thick AGN using newly reported \emph{NuSTAR} data, although the data are poor (see Table\,\ref{table:nh}). No absorption measurements are reported for Fornax\,A or NGC\,3379. Among the other eight sources, we found that four objects (M\,51a, NGC\,3982, M\,106, and NGC\,4941) show absorption above $\rm{N_{H}>10^{23}cm^{-2}}$ while three (NGC\,253, MRK\,266SW, and NGC\,4395) do not seem to be obscured. No information is found for NGC\,3607. \citet{Terashima02} analysed the X-ray ASCA spectrum of NGC\,3607 and did not find evidence that this sources could be classified as an AGN. In fact, \citet{Terlevich02} suggested that X-ray emission of this source may be linked to stellar processes \citep[see also ][]{Flohic06}. Therefore, we will discard this source. For the others, therefore, is not clear to us that obscuration might be responsible for a wrong estimation of the intrinsic luminosity of the disk. If that is ruled out, an intrinsic different accretion disk emission might be the reason of such low $\rm{\alpha_{ox}}$. Indeed, this has been argued for LLAGN for which has been proposed that the accretion disk might be intrinsically different. The most accepted model for these objects is a disk that heats and turns into an optically thin, geometrically thick inefficient accretion disk \citep[i.e. an advection dominated accretion flow (ADAF),][]{Narayan-95}.

\section{Summary and discussion}
\label{sec:discussion}
The aim of this paper is to provide a catalog of candidates for fading and rising AGN in the nearby Universe using multi-wavelength observations. For this purpose we started with the 579 and 419 AGN with [OIII] and MIR continuum fluxes from \citet{Berney15} and \citet{Asmus-15, Stern-15}, respectively (121 objects in common). In both samples, we restricted to objects with $\rm{z<0.04}$. Altogether, the initial sample contains 877 nearby AGN, all of them with available X-ray luminosity.

The hypothesis behind this multi-wavelength comparison is that each luminosity is tracing a different component: the X-ray continuum is a tracer of the disk emission, MIR is a tracer of the AGN dusty torus, and [OIII] is a tracer of the NLR emission. Since each of them occupy a different spatial scale, they might trace recent (up to $\rm{\sim}$ 3000\,yr) changes on the bolometric luminosity of the system. This allowed us to select fading or rising candidates as those out of the known linear relations between X-ray versus [OIII] and X-ray versus MIR luminosities (see Section\,\ref{sec:scalingrelations}). Using this technique we selected 161 AGN fading/rising candidates. 

We complemented [OIII], MIR, and X-ray luminosities (corrected from obscuration along the line-of sight) for 110 of these candidates to study the monotonic behaviour of the AGN activity. Among them, we discard eight sources because they were close to 2-$\rm{\sigma}$ from the linear relations in Fig.\,\ref{fig:scalingrelation}, remaining 102 candidates. We found that: 1) 53 objects are consistent with a fading scenario (called as fading); 2) five objects only show this fading scenario in the comparison between the X-ray and MIR luminosity but not when comparing MIR and [OIII] luminosities (called as early fading); 3) one object shows a monotonic increase of the bolometric luminosity between the three wavelengths (called rising); and 4) two objects show an increase of the bolometric luminosity between X-ray and MIR, but show consistent values with the linear relation between MIR and [OIII] luminosity (called early rising). We further explore more complex behaviours named atypical 1,  atypical 2, and atypical 3, with eight, 28, and five objects belonging to these categories, respectively (see Section\,\ref{sec:candidates} and Fig.\,\ref{fig:prototypes}). 

We also explored the robustness of the selection using available optical emission line diagnostics, MIR spectra, and UV continuum luminosity (see Section\,\ref{sec:robustness}). Through these comparisons, we rule out six sources belonging to the category atypical 1 because the $\rm{L [OIII]}$ could be attenuated due to dust and a proper correction of the luminosity might discard them as fading candidates. Moreover, we also exclude from our statistics the other two objects  belonging to the group atypical 1 because they do not show a consistent rising or fading scenario. We also discard NGC\,3607, due to previous evidence that the nuclear emission could be associated with stellar processes. These seven sources are marked with dagger next to the name in all the tables. Although their nature might indicate abrupt changes on the disk luminosity, due to the complex behavior of the group atypical 3 (black dotted line in Fig.\,\ref{fig:prototypes}) we do not further discuss them here. However, the group called atypical 2 is kept within the fading candidates because they might be in a late stage of the fading phase (see below). Therefore, our bonafide sample includes 88 candidates (52 fading, 28 atypical-2, five early fading, one rising, and two early rising).

\subsection{Comparison with reported candidates and caveats}

Previously reported dying AGN are Arp\,187 \citep[][]{Ichikawa18} and NGC\,7252 \citep[][]{ Schweizer13} while fading AGN are IC\,2497 \citep[][]{Jozsa09}, the Teacup Galaxy \citep[also known as SDSS\,J143029.88+133912.0][]{Keel12}, Mkn\,1498, NGC\,5252, NGC\,5972, SDSS\,J151004.01+074037.1, SDSS\,J220141.64+115124.3, UGC\,7342, and UGC\,11185 \citep[][]{Keel17}. We find that three (namely Mkn\,1498, UGC\,11185, and NGC\,5252) out of these 11 sources were included initially among our samples in Section\,\ref{sec:scalingrelations}. However, none of the three are classified as fading AGN according to our criterion. They are selected by \citet{Keel12} for having ionized cones of over 10\,kpc; therefore, all of them show extended NLR. Thus, the first caveat in this methodology is that we might be missing fading type-2 AGN because the NLR emission is too extended to be included in the [OIII] fluxes presented in this work. However, note that 38 of the fading candidates are type-2 AGN (47\% of the sample) while 42 are type-1 AGN (53\%), as reported in the last column in Tables\,\ref{tab:fluxes} and \ref{tab:fluxes2}. Thus, we are not missing all type-2 AGN but probably only those with a very extended NLR.

Another caveat of this classification is that the accretion disk or the NLR signatures might be obscured (and not properly corrected) in our X-ray or [OIII] luminosity. However, we have used [OIV] emission line luminosity finding that, apart from the so-called atypical 1 class where the [OIII] might be attenuated, [OIII] and [OIV] emission lines seem to behave as expected in AGN (see Fig.\,\ref{fig:LOIIIvsLOIV}).

Furthermore, we also explored X-ray versus UV continuum correlations to search for candidates highly obscured at X-rays. The obtained correlations of $\rm{ log(L_\nu(\lambda2500}$\,\AA)) versus $\rm{log(L_\nu(2keV))}$ and $\rm{\alpha_{ox}} versus\,  \rm{log(L_{\nu}(2500}$\,\AA)) are in agreement with previous work \citep{just07,xu11}. Thus, we rule out a wrong estimate of the intrinsic disk luminosity for the overall sample. However, the scatter of the sources is obvious and non negligible, with four extreme objects having $\rm{\alpha_{ox}} <$ -2.5 and 11 sources having $-2.5  < \rm{\alpha_{ox}} < -1.5$. However, even for those sources, we did not find indications of a wrong estimate of the $\rm{N_H}$. Fig.\,\ref{fig:histograma_accepted} shows the distribution of $\rm{N_H}$ when available for fading/rising AGN (left) and atypical AGN (right). Most of our sources are classified as obscured sources at X-rays (i.e. $\rm{N_H>10^{22}cm^{-2}}$). These measurements are taken from \emph{Swift}/BAT or \emph{NuSTAR} observations, ensuring that, even for Compton-thick AGN (i.e. with $\rm{N_H > 3\times 10^{24}\, cm^{-2}}$), the X-ray luminosity is properly corrected from this attenuation. 

We are capable of measuring recent (last $\rm{\sim}$3000\,yr for fading/rising and $\rm{\sim}$30\,yr for early fading/rising) luminosity changes. Another caveat to bare in mind is that large amplitude disk continuum variations might result in similar rise/fade of the nuclear signatures compared to the luminosity of the NLR or the dusty structure, as the ones presented here. However, in particular for the fading and rising candidates, this change must be sustained for a long period of time to produce a consistently decreasing/increasing behaviour on their luminosity. Indeed, in changing-look AGN with such long term disk-related changes has long been discussed \citep{Matt03}. Despite systematic search for changing look QSOs candidates at distance beyond our parent samples \citep[][]{Graham17,Rumbaugh18,MacLeod19}, a few dozen Seyfert galaxies are known to have changed their optical spectral type. Here we present a compilation of objects belonging to this category: NGC\,3516 \citep[][]{Collin-Souffrin73}, NGC\,7603 \citep[][]{Tohline76, Kollatschny00}, NGC\,4151 \citep[][]{Penston84}, Fairall\,9 \citep[][]{Kollatschny85}, NGC\,2617 \citep[][]{Shappee14}, Mrk\,590 \citep[][]{Denney14}, HE\,1136-2304 \citep[][]{Parker16}, 1ES\,1927+654 \citep[][]{Trakhtenbrot19}, IRAS\,23226-3843 \citep[][]{Kollatschny20}, SDSSJ\,095209.56+214313.3 \citep[][]{Komossa08}, Mrk\,1018 \citep[][]{Noda18}, and ESO\,121-G006 \citep{Annuar20}. Interestingly, only NGC\,4151 is classified here as a fading candidate while seven (Fairall\,9, NGC\,3516, NGC\,4151, NGC\,7613, HE\,1136-2304, Mrk\,590, and Mrk\,1018) out of these 11 were included within the 877 sources analysed in the initial sample of this paper. Based on dynamical, thermal, and viscous time-scales, \citet{Ichikawa19} suggested that the luminosity changes in dying and changing-look AGN are likely based on the different physical mechanisms of the accretion disk. While changing-look AGN might be associated with thermal timescales corresponding to the disk cooling, dying AGN are more likely associated with the viscous timescale of the accretion process. To investigate if some of these candidates are indeed changing look AGN or persistent fading/rising of their AGN activity, follow-up observations are needed. This can allow us to study the long-term variations in order to try to characterize the plausible disappearance of broad lines within a few years through optical spectroscopy \citep[as for instance][]{Lawrence18}.

\vspace{0.5cm}
\subsection{AGN components}

Additional support of the fading stage of these sources comes from the analysis of the AGN components. Here we explore the AGN dust and jet. To study the AGN dust, we compiled \emph{Spitzer}/IRS spectra available for 31 AGN discussed here\footnote{We compiled 38 \emph{Spitzer}/IRS spectra, 31 among them are classified within the fading/rising and the atypical-2 classes.} (see Tables\,\ref{tab:fluxes} and \ref{tab:fluxes2}). We decompose the spectra into AGN dust and circumnuclear contributors (i.e., stellar and interstellar medium). For the AGN dust we used a set of five available models in the literature (see Appendix\,\ref{app:SpectralFit} for more details). Among the 22 objects where the AGN dust dominates (15 classified as fading, one as early fading, and six as atypical 2) the torus-like morphology is preferred against the disk-wind in our sample, with only between 3 objects preferring the latter (five if we add two that are equally fitted with either a torus-like geometry or the disk+wind model). This result is opposite to that found for nearby AGN by \citet{Gonzalez-Martin19B}, where the largest percentage of good fits is obtained for the clumpy disk+wind model by \citet{Hoenig17}. 

AGN radio lobes are also analysed to look for the AGN aging, through the study of the kinematic jet age of the radio lobes \citep[e.g.][]{Ichikawa19}. Although, we leave the kinematic jet age estimates for a subsequent analysis, we explore here if these jets, as long standing signatures of past activity, are present within our sample of candidates. We found that among all the sources with available radio images (a total of 61 available radio images and 55 within the secure sample of fading/rising of activity), the percentage of clear \textit{Linear} sources is $\sim 31\%$. We characterized the radio morphology of fading, rising and atypical objects (see Col. 9 in Tables \,\ref{table:candidates} and \ref{table:rejected}) after looking for radio images in the literature (see Appendix\,\ref{sec:RadioImages} for more details). This percentage is very large considering that the percentage of local AGN showing powerful radio jet has been measured to be of the order of 0.1\% \citep[rising up to 10\%, for high redshift quasars,][]{Blandford2019}. Moreover, considering that half of the \textit{Compact} sources could show, after performing deeper and more sensitive radio observations, a \textit{Linear} morphology (hence, showing the presence of a radio jet), this percentage of radio jets within our sample should be considered as a lower limit. This suggests that relativistic jets are produced in the centre of a large portion of these (mostly) AGN fading candidates.

Both the lack of outflows and the presence of radio jets are expected in the fading phase of nuclear activity, at least in the context of BH X-ray binaries. When they enter into a burst, they evolve into two distinct states along their duty cycle, known as hard and soft states. The hard state occurs at the beginning and at the end of the burst, while the soft state is associated with the most efficient accretion rate along the burst. In the soft state there is a weak or non-existent core jet and a strong accretion disk wind. On the other hand, the hard state is associated with a powerful, quasi-steady jet linked to the initial rise and fade of the transient event \citep[][and references therein]{Fender12,Fender16}. Therefore, both in the rising or fading stages of the duty cycle, jets are expected and winds/outflows are lacking.

Further support of the lack of winds and the launch of jets within the low accretion AGN state comes from theory. During this stage, SMBHs may switch to a different accretion mode, characterized by a corona of low accretion rate and low-radiative efficiency \citep[radiatively inefficient accretion flows models, RIAFs][]{Narayan05}. Thus, the geometrically thin and optically thick disk might not be present along this AGN stage. The funnels in the geometrically thick toroidal-RIAFs are invoked as a plausible mechanism for collimating the jet \citep[][]{Nagar02}. Furthermore, under the wind model, a minimal accretion rate is required to produce a minimal column density to detect the outflows \citep[][]{Elitzur16}. Therefore, the lack of dusty winds in our sample might be the manifestation of the lack of accretion power to sustain the wind/outflow activity. Note, however, that our sample covers a wide range in X-ray luminosity, indicating that this might not be as simple as a luminosity threshold. Indeed, theoretical studies show that the existence of these winds might rely on the particular configuration of the wind \citep{Elitzur09,Elitzur16}. Our results are consistent with this framework.

\subsection{Implications on the duty cycle of AGN}

Altogether, we found 85 fading candidates (including fading, early fading, and atypical 2) and 3 rising candidates among the 877 nearby analyzed sources. Most of them are Seyfert nuclei with only 8 LINERs (see Tables\,\ref{tab:fluxes} and \ref{tab:fluxes2}). Thus, we find $\rm{\sim 10\%}$ ($\rm{\sim 0.3\%}$) of fading (rising) AGN in the local Universe. Statistically speaking, this might imply that one tenth of the AGN duty cycle of activity ($\rm{\sim}$10\,Myr) is spent in this fading phase \citep[][]{Hopkins05}. This cycle can be explained as the activation of the nuclear accretion towards the SMBH due to a merger process or strong disk instabilities. Interestingly, roughly half of the sample (40 of these 88 sources) are associated with pairs, groups, or systems with multiple objects (e.g. galaxy clusters). Thus, we would expect to see a prevalence for elliptical galaxies. However, among these 41 sources, we do not see a preference for any morphological type (12 elliptical/S0/irregulars and 28 spiral galaxies).

Under the merger event, the gas supplied will be 100 times larger than that needed to efficiently feed the SMBH. Thus, the rising phase is expected to happen quickly. This might explain the very few objects found in the rising phase in this analysis. Then, the SMBH growth continues while gas suppliers start to skimp, until the gas is depleted and the duty cycle ends \citep{Hopkins12}. Thus, the fading phase is expected to last longer than the rising phase, as statistically found in this paper. 

Alternatively, the lack of rising AGN in our sample, might be a natural consequence of the switch off of the star-formation and AGN activity at the present time as a result of gas-rich mergers occurred in the past \citep{Hopkins08}. The star-formation rate density peaked approximately 3.5\,Gyr after the Big Bang, at z$\rm{\sim 2}$ (known as cosmic noon), and declined exponentially at later times \citep[see][for a review]{Madau14}. Studies at higher redshift might help to corroborate this hypothesis studying the fraction of rising and fading AGN at the peak of the star-forming activity.

\begin{figure}
\begin{center}
\includegraphics[width=1.0\columnwidth]{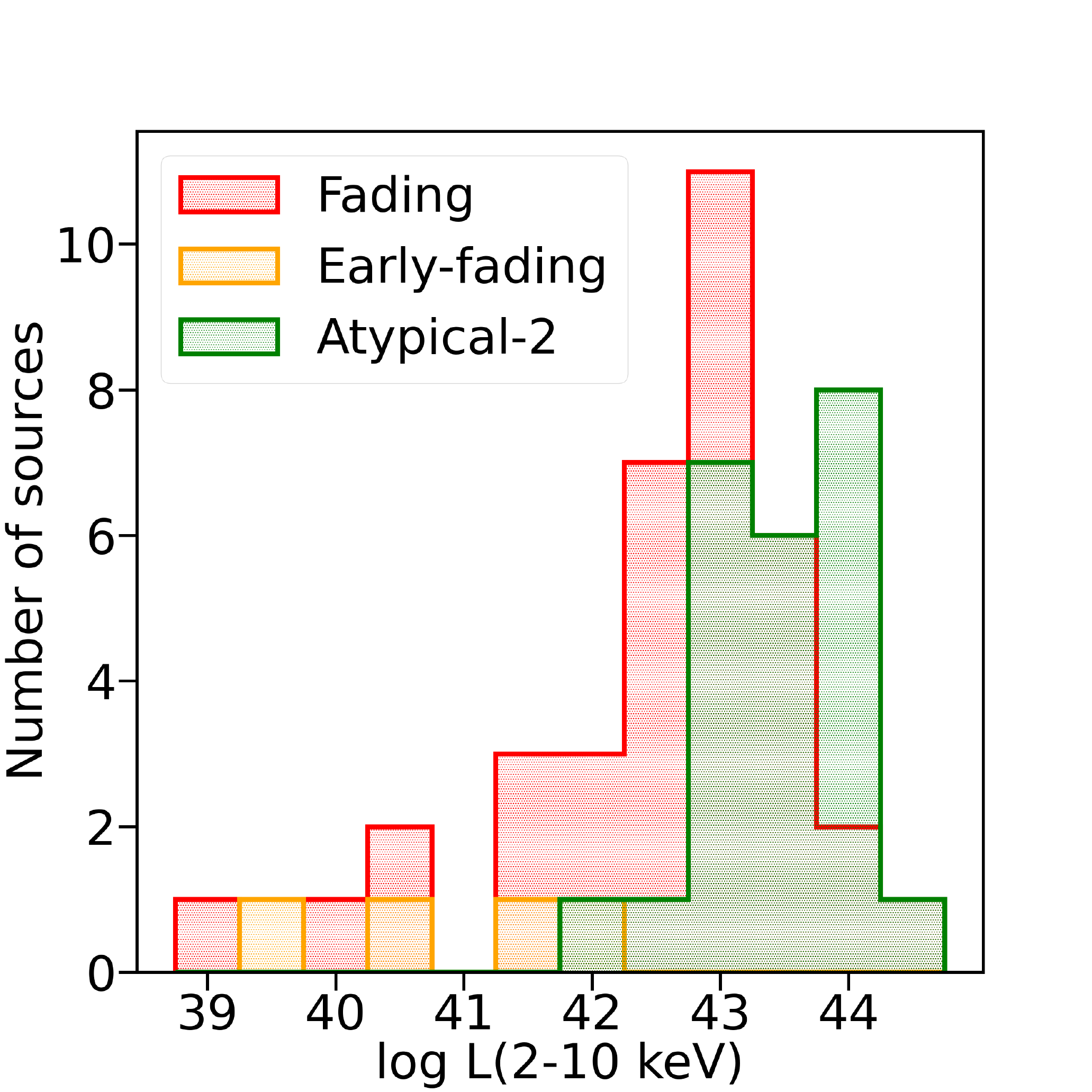}
\caption{Histogram of 2-10 keV X-ray luminosity, $\rm{L(2-10 keV)}$ for the fading, early fading and atypical 2 candidates.}
\label{fig:Lxhist}
\end{center}
\end{figure}

If the fading phase is long enough we might expect to see different stages along it. Interestingly, we could see at least three categories: early fading, fading, and atypical 2. Most of these sources are nicely classified as AGN in the BPT diagram (see Fig.\,\ref{fig:BPT}), placed in the [OIV] versus [OIII] emission line luminosity relation (see Fig.\,\ref{fig:LOIIIvsLOIV}), neglecting extinction to be affecting these three classes. The fading AGN category contains 52 AGN showing a monotonically increase of AGN bolometric luminosity, with a lower value for the disk and larger values for the torus and the NLR. 
Five objects belong to the early fading class, showing disk bolometric luminosity lower than that inferred from the torus/NLR but with the torus and the NLR with compatible bolometric luminosity. We think that in these five sources the fade of the nuclear activity is more recent than that of AGN fading candidates so the outer parts of the AGN have not adjusted yet to the current fading of the AGN accretion disk (named here early fading). Moreover, 28 objects belonging to the category atypical 2 are consistent with a late fading phase, where the disk and torus are consistent with the same bolometric luminosity but the NLR still reflects a larger bolometric luminosity.

Interestingly, these three classes of sources show slightly different average bolometric luminosity of the disk. To illustrate this, Fig.\,\ref{fig:Lxhist} shows the histogram of X-ray luminosity for these three categories. The early fading, fading, and atypical 2 (late fading) classes show different average disk luminosity, where atypical 2 show the lowest luminosity, fading candidates show intermediate luminosity, and early fading candidates show the highest. This is also consistent with the idea that these three classes are indeed stages of the same evolutionary track towards the AGN activity switch off.

\section{Conclusions}
\label{sec:conclusions}

We have found a bonafide sample of 88 candidates (out of 877 AGN) of rising or fading of the AGN activity in the nearby Universe using AGN scaling relations. In particular, we explored the MIR versus X-ray and the [OIII] versus X-ray relations. We selected fading/rising candidates as those out of these relations. We also used multi-wavelength information (mainly optical BPT, the UV luminosity, and the MIR spectrum) to explore sources contaminated by extra-nuclear emission, X-ray luminosity not well corrected from absorption in highly obscured AGN, and non AGN powered sources.

Around 10\% of our initial sample presents a fading/rising scenario. Furthermore, the vast majority of these candidates ($\rm{\sim 96\%}$) are fading sources. This may be explained if the Universe had its peak of activity back in the past, currently dominated by the AGN activity switch off. Alternatively, this might indicate that the fading phase is longer than the rising phase. The current sample of fading candidates might be missing type-2 AGN because the NLR emission is extended. The large amplitude disk continuum variations might also result in similar rise/fade of the nuclear signatures compared to the luminosity of the NLR or the dusty structure, perhaps including some changing look AGN. Follow-up observations are needed in order to confirm these candidates.

We found that these (mostly) fading candidates are placed within merging or interacting systems. We also found that among our sample, $\sim 31\%$ of our AGN had a clear evidence for the existence of a linear radio source (jet) which is higher than that estimated in the nearby Universe. Moreover, we also found a prevalence of AGN dust associated with torus-like geometries rather than outflows. The lack of outflows and the presence of radio-jets are expected in the fading phase of nuclear activity. In fact, \citet{King11} suggests that large-scale outflows may persist for as long as 100 Myr after a powerful AGN episode fades \citep[see also][]{Zubovas14}.

\section*{Acknowledgments}

We thank to the anonymous referee for his/her useful comments. This research has made use of the NASA/IPAC Extragalactic Database (NED) which is operated by the Jet Propulsion Laboratory, California Institute of Technology, under contract with the National Aeronautics and Space Administration. This work is based in part on observations made with the \emph{Spitzer} Space Telescope, which is operated by the Jet Propulsion Laboratory, California Institute of Technology under a contract with NASA. This research is mainly funded by the UNAM PAPIIT project IA103118 (PI OG-M). DE-A, NO-C, CV-C, and UR-A acknowledge support from a CONACYT scholarship.

\begin{deluxetable*}{lcccccccccccc}
\tablecolumns{13}
\tablewidth{10cm}
\tablecaption{Final sample of rising/fading candidates\label{table:candidates}}
\tabletypesize{\tiny}
\tablehead{
\colhead{Name} & \colhead{RA} & \colhead{Dec} & \colhead{z} & \colhead{$\rm{\nu L (12\mu m)}$} & \colhead{$\rm{L_{X}}$} & \colhead{$\rm{L_{[OIII]}}$} & \colhead{candidate}  & \colhead{radio}  & \colhead{$\rm{N_H}$} & \colhead{$\rm{log(\nu L_{\nu})}$} & \colhead{hubty} & \colhead{env} \\
\vspace{-0.2cm}
 &  &  &   &   &   &   & \colhead{class} & \colhead{morph.} & $\rm{cm^{-2}}$ &\colhead{\texorpdfstring{$\lambda$}{lambda-}2500 \AA} &  & \\
\hspace{1cm} (1) & (2) & (3) & (4) & (5)  & (6)  & (7)  & (8)  & (9) & (10) & (11) & (12) & (13)
}
\startdata
Mrk\,335 & 00h06m19.52s & 20d12m10.5s & 0.0351 & 44.34 & 43.5 & 41.63 & Fading & Compact & 20.48 & 44.27  & E &-- \\
J00430184+3017195 & 00h43m01.87s & 30d17m19.6s & 0.0441 & 44.08 & 43.07 & 41.77 & Fading & -- & 22.30& 41.86    &-- &-- \\
Mrk\,359 & 01h27m32.55s & 19d10m43.8s & 0.0167 & 43.29 & 42.68 & 40.89 & Fading & Compact & 20.61& 43.04   & S0 & group\\
NGC\,1068 & 02h42m40.71s & -00d00m47.8s & 0.0032 & 43.79 & 42.82 & 41.72 & Fading & Linear$^4$ & 24.95& 43.07   & Sb & group\\
Fornax\,A & 03h22m41.72s & -37d12m29.6s & 0.0043 & 41.26 & 40.3 & 38.73 & Fading & Diffuse$^3$  & -- & 42.67   & S0 & pair\\
IRAS\,04124-0803 & 04h14m52.67s & -07d55m39.9s & 0.0382 & 44.21 & 43.27 & 42.22 & Fading & -- & -- & 43.09    &-- &-- \\
Mrk\,618 & 04h36m22.24s & -10d22m33.8s & 0.0356 & 44.39 & 43.4 & 41.73 & Fading & Compact & -- & 43.76    & SBb &--\\
LEDA\,097068 & 05h02m58.22s & 22d59m51.8s & 0.0577 & 44.79 & 43.98 & 42.50 & Fading & -- & -- & 44.27    & E &--\\
IRAS\,05218-1212 & 05h24m06.50s & -12d09m59.6s & 0.049 & 44.26 & 43.39 & 42.56 & Fading & -- & -- &43.44    & E &--\\
Mrk\,6 & 06h52m12.25s & 74d25m37.5s & 0.0222 & 43.81 & 43.25 & 42.08 & Fading & Linear$^6$ & 20.76& 42.62   & S0-a & pair\\
Mrk\,79 & 07h42m32.80s & 49d48m34.7s & 0.0316 & 44.25 & 43.43 & 41.79 & Fading & Linear$^6$& -- & 43.75   & Sb & -- \\
Mrk\,10 & 07h47m29.13s & 60d56m00.6s & 0.0292 & 43.84 & 43.13 & 41.53 & Fading & Compact & 20.53& 44.16    & SABb &--\\
Mrk\,1210 & 08h04m05.86s & 05d06m49.8s & 0.0135 & 43.72 & 43.15 & 41.52 & Fading & Compact &23.40 & 42.31   & S? &--\\
IRAS\,09149-6206 & 09h16m09.39s & -62d19m29.9s & 0.0573 & 44.96 & 43.95 & 42.59 & Fading & -- & ${24.19}_{-0.05}^{+0.05}$* & 44.74    &-- &--\\
Mrk\,704 & 09h18m26.01s & 16d18m19.2s & 0.0292 & 44.25 & 43.4 & 41.83 & Fading & Compact & ${22.72}_{-0.07}^{+0.51}$* & 43.63    & S0-a & group \\
M\,81 & 09h55m33.17s & 69d03m55.1s & 0.0008 & 41.50 & 38.8 & 38.64 & Fading & Diffuse$^1$& ${23.58}_{-0.03}^{+0.03}$* & 39.94   & Sab & triple\\
3C\,234.0 & 10h01m49.52s & 28d47m09.0s & 0.1849 & 45.50 & 44.41 & 43.19 & Fading & Linear$^7$  & 23.51& 43.98   &--  &--\\
NGC\,3227 & 10h23m30.58s & 19d51m54.2s & 0.0043 & 43.09 & 42.22 & 40.59 & Fading & Diffuse$^1$ &20.95 & 41.03   & SABa & pair\\
ESO\,317-G038 & 10h29m45.61s & -38d20m54.8s & 0.0151 & 43.20 & 41.58 & 40.80 & Fading & -- &23.41 & 42.37    &SBa &--\\
NGC\,3379 & 10h47m49.59s & 12d34m53.8s & 0.003 & 40.22 & 38.29 & 38.15 & Fading & Compact & -- & 41.68    & E & triple\\
NGC\,3521 & 11h05m48.58s & -00d02m09.1s & 0.0027 & 40.92 & 38.20 & 38.14 & Fading & Linear$^8$ & -- & 42.71   & SABb & -\\
ESO\,438-G009 & 11h10m48.00s & -28d30m03.8s & 0.0219 & 43.95 & 42.65 & 40.99 & Fading & -- & ${24.51}_{-0.31}^{+0.46}$* & 43.83    &SBab &--\\
NGC\,3607$^{\dagger}$ & 11h16m54.64s & 18d03m06.3s & 0.0032 & 40.65 & 38.73 & 39.12 & Fading & Compact & -- & 41.28    & E-S0 & pair\\
PG\,1138+222 & 11h41m16.16s & 21d56m21.8s & 0.0632 & 44.44 & 43.80 & 42.29 & Fading & Compact & -- &43.40    & Sab &--\\
NGC\,3982 & 11h56m28.13s & 55d07m30.9s & 0.0048 & 41.55 & 40.2 & 40.03 & Fading & Compact & ${23.83}_{-0.20}^{+0.18}$* & 41.90    &SABb & group\\
UGC\,07064 & 12h04m43.32s & 31d10m38.2s & 0.0247 & 43.78 & 42.59 & 41.26 & Fading & Diffuse$^2$ & 22.59& 42.54   & SBb & triple\\
NGC\,4151 & 12h10m32.58s & 39d24m20.6s & 0.0023 & 42.58 & 42.03 & 41.28 & Fading & Linear$^4$ &22.71 & 42.18   & Sab & pair\\
Mrk\,766 & 12h18m26.51s & 29d48m46.3s & 0.0129 & 43.57 & 42.73 & 41.23 & Fading & Compact &20.32 & 42.43    &SBa & pair\\
NGC\,4303 & 12h21m54.90s & 04d28m25.1s & 0.0052 & 38.38 & 36.52 & 38.73 & Fading & Diffuse$^1$ & -- & 43.96   & Sbc & pair\\
NGC\,4395 & 12h25m48.86s & 33d32m48.9s & 0.0009 & 40.87 & 40.43 & 38.91 & Fading & Diffuse$^1$ &21.04 & 42.39   & Sm & multiple\\
J123212.3-421745 & 12h32m11.83s & -42d17m52.2s & 0.1009 & 44.85 & 44.12 & 42.74 & Fading & -- &  -- & 44.46    &-- &--\\
LEDA\,170194 & 12h39m06.29s & -16d10m47.1s & 0.0360 & 43.54 & 42.95 & 42.23 & Fading & -- &22.76 & 43.14    & S0 &--\\
NGC\,4736 & 12h50m53.06s & 41d07m13.6s & 0.0011 & 39.95 & 38.56 & 37.45 & Fading & Diffuse$^1$ & -- & 42.37   & SABa & group\\
NGC\,4748 & 12h52m12.46s & -13d24m53.0s & 0.0142 & 43.30 & 42.34 & 41.36 & Fading & Compact & -- & 42.87    & S? &--\\
NGC\,4941 & 13h04m13.14s & -05d33m05.8s & 0.0040 & 42.50 & 41.28 & 40.09 & Fading & Linear$^6$ &23.72 & 43.11   & SABa & group\\
NGC\,4939 & 13h04m14.39s & -10d20m22.6s & 0.0105 & 43.24 & 42.38 & 41.08 & Fading & Diffuse$^1$ & 23.29& 41.46   & Sbc & group\\
MCG-03-34-064 & 13h22m24.46s & -16d43m42.5s & 0.0199 & 44.34 & 43.59 & 41.80 & Fading & Linear$^6$ & 23.80& 42.59   & S0 &\\
M\,51a & 13h29m52.71s & 47d11m42.6s & 0.0018 & 40.69 & 40.4 & 38.94 & Fading & Diffuse$^1$& ${24.67}_{-0.06}^{+0.06}$* & 43.01   & SABb & pair\\
ESO\,509-G038 & 13h31m13.90s & -25d24m10.0s & 0.0263 & 43.78 & 42.51 & 41.48 & Fading& -- & ${23.90}_{-0.06}^{+0.07}$* & 43.02    & S0-a &--\\
NGC\,5283 & 13h41m05.76s & 67d40m20.3s & 0.0103 & 42.73 & 41.95 & 40.96 & Fading & Linear$^4$ &23.15 & 41.74   & S0 &--\\
NGC\,5273 & 13h42m08.34s & 35d39m15.2s & 0.0038 & 42.02 & 41.37 & 39.49 & Fading & Compact & 20.59& 40.93    & S0 & pair\\
Mrk\,463 & 13h56m02.87s & 18d22m19.5s & 0.0503 & 44.88 & 43.10 & 42.44 & Fading & Linear$^9$ &23.57 & 43.74   & Sc &--\\
Mrk\,477 & 14h40m38.10s & 53d30m15.9s & 0.0377 & 43.97 & 43.26 & 42.50 & Fading & Compact & 23.52& 43.51    & S0 & pair \\
IC\,4518A & 14h57m41.18s & -43d07m55.6s & 0.0163 & 43.57 & 42.66 & 41.22 & Fading & -- &23.36 & 44.00    & Sc &  multiple\\
Mrk\,1392 & 15h05m56.55s & 03d42m26.3s & 0.0356 & 43.98 & 43.10 & 41.81 & Fading & Compact & ${24.63}_{-0.18}^{+0.16}$*  & 43.70    & SBcd & group\\
J15462424+6929102 & 15h46m24.33s & 69d29m10.0s & 0.0376 & 43.72 & 43.08 & 41.86 & Fading & Linear$^2$ &23.49 & 42.33   &--  &--\\
J16531506+2349431 & 16h53m15.05s & 23d49m43.0s & 0.1031 & 44.80 & 44.03 & 42.65 & Fading & Compact & 23.27& 43.11    & E &--\\
Fairall\,49 & 18h36m58.29s & -59d24m08.6s & 0.0201 & 44.17 & 43.40 & 41.37 & Fading & -- &22.03 & 42.34    & E-S0 &--\\
J19373299-0613046 & 19h37m33.01s & -06d13m04.8s & 0.0103 & 43.48 & 42.77 & 41.44 & Fading & -- & 20.85& 43.32    & E &--\\
MCG+02-57-002 & 22h23m45.02s & 11d50m09.0s & 0.0294 & 43.31 & 42.63 & 41.64 & Fading & Diffuse$^2$ & -- & 43.86   & Sb &--\\
Mrk\,915 & 22h36m46.50s & -12d32m42.6s & 0.0239 & 43.77 & 43.20 & 42.00 &  Fading & Compact & ${23.53}_{-0.09}^{+0.07}$* & 43.79    & Scd & triple\\
MCG+01-57-016 & 22h40m17.05s & 08d03m14.1s & 0.0249 & 43.87 & 43.04 & 41.78 & Fading & Compact & -- & 43.72    & SBa &--\\
NGC\,7469 & 23h03m15.62s & 08d52m26.4s & 0.0139 & 43.90 & 43.2 & 41.82 & Fading & Diffuse$^4$ & 20.53& 43.13   & Sa & pair \\\hline 
NGC\,3627 & 11h20m14.96s & 12d59m29.5s & 0.0024 & 40.98 & 39.5 & 37.84 & Early Fading &  Diffuse$^1$ & ${24.26}_{-0.45}^{+0.67}$* & 43.25   & Sb & pair\\
NGC\,4051 & 12h03m09.61s & 44d31m52.8s & 0.0031 & 42.89 & 41.63 & 39.91 & Early Fading &  Diffuse$^1$ & ${24.53}_{-0.02}^{+0.02}$* & 42.14   & SABb & group\\
M\,106 & 12h18m57.50s & 47d18m14.3s & 0.0016 & 42.53 & 40.61 & 39.08 & Early Fading &  Diffuse$^5$ & 23.00& 43.10   & Sbc & pair\\
NGC\,5033 & 13h13m27.47s & 36d35m38.2s & 0.0028 & 42.76 & 40.91 & 39.25 & Early Fading & Diffuse$^1$ & -- & 42.18   & Sc & group\\
NGC\,7130 & 21h48m19.52s & -34d57m04.5s & 0.0161 & 43.95 & 42.15 & 40.95 & Early Fading & Compact & 24.00& 43.12    & Sa & pair\\\hline
NGC\,1194 & 03h03m49.11s & -01d06m13.5s & 0.0131 & 43.44 & 43.67 & 39.97 & Early Rising &  Linear$^6$ & 24.33& 40.98   & S0a & multiple \\
J14391186+1415215 & 14h39m11.87s & 14d15m22.0s & 0.0717 & 43.45 & 43.74 & 40.00 & Early Rising &  -- & 22.40& 42.06    & E &--\\\hline
J08551746-2854218 & 08h55m17.47s & -28d54m21.4s & 0.073 & 43.64 & 43.80 & 39.96 & Rising & -- &21.95 & 43.87    &-- &-- \\
\enddata
\tablecomments{\scriptsize The sources removed in section \ref{sec:discussion} are marked with a dagger next to the name in Col.(1). Asterisks alongside the value represent those objects with a wrong column density estimate, for which a further analysis was carried on (see Appendix A1), Dashes refer to as those objects with no information. Col (11). Logarithmic of luminosity (erg/s) at \texorpdfstring{$\lambda$}{lambda-}2500 \AA. Data from the main search engine the Nasa Extragalactic Database (NED). References: 1-\cite{Condon(1987)}; 2-\cite{Smith2016}; 3-\citet{Fomalont1989}; 4-\cite{Kukula1995}; 5-\cite{Hummel1985}; 6-\cite{Schmitt2001};7-\cite{Hardcastle1997};8-\cite{Hummel1987};9-\cite{Ulvestad1981}.}
\end{deluxetable*}

\begin{deluxetable*}{lcccccccccccc}
\tablecolumns{13}
\tablewidth{10cm}
\tablecaption{Sample of atypical candidates.\label{table:rejected}}
\tabletypesize{\tiny}
\tablehead{
\colhead{Name} & \colhead{RA} & \colhead{Dec} & \colhead{z} & \colhead{$\rm{\nu L (12\mu m)}$} & \colhead{$\rm{L_{X}}$} & \colhead{$\rm{L_{[OIII]}}$} & \colhead{candidate}  & \colhead{radio}  & \colhead{$\rm{N_H}$} & \colhead{$\rm{log(\nu L_{\nu})}$} & \colhead{hubty} & \colhead{env} \\
\vspace{-0.2cm}
 &  &  &   &   &   &   & \colhead{class} & \colhead{morph.} & $\rm{cm^{-2}}$ &\colhead{\texorpdfstring{$\lambda$}{lambda-}2500 \AA} &  & \\
\hspace{1cm} (1) & (2) & (3) & (4) & (5)  & (6)  & (7)  & (8)  & (9) & (10) & (11) & (12) & (13)
}
\startdata
NGC\,612$^{\dagger}$ & 01h33m57.74s & -36d29m35.7s & 0.0301 &  44.02 & 43.94 & 40.08 & Atypical 1 & Linear $^1$  & 23.99 & 42.79 & S0-a & multiple\\
J02420381+0510061$^{\dagger}$ & 02h42m03.80s & 05d10m06.1s & 0.0711 &  44.25 & 43.56 & 40.21 & Atypical 1 & --  & 23.50 & 42.58 & -- & -- \\
J04440903+2813003$^{\dagger}$ & 04h44m09.01s & 28d13m00.7s & 0.0107 & 43.22 & 42.64 & 38.74 & Atypical 1 & --  & 22.65 & 43.50 & Sb & -- \\
PKS\,0558-504$^{\dagger}$ & 05h59m47.38s & -50d26m52.4s & 0.1372 &  45.04 & 44.85 & 41.26 & Atypical 1 & --  & ${25.46}_{-0.50}^{+1.00}$* & 45.20 & -- & -- \\
NGC\,3079$^{\dagger}$ & 10h01m57.80s & 55d40m47.2s & 0.0036 &  43.29 & 41.30 & 37.54 & Atypical 1 & Diffuse$^2$ & 25.10 & 42.66 & SBc & pair \\
Cen\,A$^{\dagger}$ & 13h25m27.62s & -43d01m08.8s & 0.00086 & 42.82 & 41.79 & 37.29 & Atypical 1 & Linear$^3$ & 23.02 & 42.20 & S0 & pair  \\
ESO\,097-G013$^{\dagger}$ & 14h13m09.950s & -65d20m21.20s & 0.00094 & 42.64 & 42.31 & 38.80 & Atypical 1 & -- & 24.40 & 42.10 & Sb & pair\\
MCG+04-48-002$^{\dagger}$ & 20h28m35.06s & 25d44m00.0s & 0.0139 & 43.77 & 43.16 & 38.78 & Atypical 1 & --  & 23.86 & 42.73 & Sd & pair\\\hline
MCG-07-03-007 & 01h05m26.82s & -42d12m58.3s & 0.0302 & 43.75 & 43.47 & 41.36 & Atypical 2 & --  & 24.18 & 42.21 & Sa & -- \\
MCG+08-03-018 & 01h22m34.43s & 50d03m18.0s & 0.0202 & 43.66 & 43.98 & 42.04 & Atypical 2 & --  & 24.24 & 41.68 & Sc & -- \\
NGC\,526A & 01h23m54.39s & -35d03m55.9s & 0.0188 & 43.46 & 43.27 & 41.68 & Atypical 2 & --  & 22.01 & 42.45 & S0 &  multiple\\
NGC\,1229 & 03h08m10.79s & -22d57m38.9s & 0.0357 & 43.42 & 43.96 & 41.54 & Atypical 2 & -- & 24.94 & 44.25 & SBbc & multiple \\
J03305218+0538253 & 03h30m52.18s & 05d38m25.6s & 0.046 & 43.94 & 43.62 & 42.46 & Atypical 2 & Compact & -- & 43.38 & -- & -- \\
CGCG\,420-015 & 04h53m25.75s & 04d03m41.7s & 0.0294 & 44.11 & 44.01 & 41.86 & Atypical 2 & Linear$^4$ & 24.14 & 43.69 & E & -- \\
Mrk\,3 & 06h15m36.36s & 71d02m15.1s & 0.0132 & 43.89 & 43.67 & 42.44 & Atypical 2 & Linear$^5$  & 24.07 & 42.10 & S0 & pair \\
Mrk\,78 & 07h42m41.73s & 65d10m37.5s & 0.037 & 44.01 & 43.82 & 42.08 & Atypical 2 & Linear$^6$& 24.11 & 42.51 & -- & --  \\
J09172716-6456271 & 09h17m27.21s & -64d56m27.1s & 0.0859 & 43.94 & 43.91 & 42.38 & Atypical 2 & --  & 21.41 & 44.02 & -- & -- \\
ESO\,374-G044 & 10h13m19.91s & -35d58m57.7s & 0.0284 & 43.95 & 43.47 & 41.65 & Atypical 2 & --  & 23.71 & 43.83 & Sab & -- \\
NGC\,3393 & 10h48m23.46s & -25d09m43.4s & 0.0138 & 42.87 & 42.74 & 41.98 & Atypical 2 & Linear$^4$ & 24.50 & 42.08 & SBa & group \\
ESO\,265-G023 & 11h20m48.01s & -43d15m50.4s & 0.0565 & 44.40 & 43.84 & 42.09 & Atypical 2 & -- & -- & 44.26 & E & pair \\
Mrk\,1310 & 12h01m14.36s & -03d40m41.1s & 0.0191 & 42.65 & 42.72 & 40.98 & Atypical 2 & -- & < 20* & 42.91 & E & -- \\
Mrk\,205 & 12h21m44.22s & 75d18m38.8s & 0.0708 & 44.13 & 43.90 & 42.47 & Atypical 2 & Compact   & ${23.69}_{-0.13}^{+0.09}$* & 44.53 & -- & -- \\
J12313717-4758019 & 12h31m37.16s & -47d58m02.0s & 0.028 & 43.64 & 43.15 & 41.43 & Atypical 2 & --  & 20.59 & 43.64 & SABb & -- \\
NGC\,4507 & 12h35m36.63s & -39d54m33.3s & 0.0117 & 43.62 & 43.53 & 41.95 & Atypical 2 &  -- & 23.95 & 43.59 & Sab & group \\
ESO\,323-32 & 12h53m20.32s & -41d38m08.3s & 0.0160 & 42.96 & 43.18 & 40.90 & Atypical 2 & --  & 24.79 & 43.86 & S0-a & group\\
Mrk\,783 & 13h02m58.84s & 16d24m27.5s & 0.067 & 44.42 & 44.01 & 42.35 & Atypical 2 & Compact & 20.78 & 44.10 & E & -- \\
NGC\,5135 & 13h25m44.06s & -29d50m01.2s & 0.0148 & 43.22 & 43.22 & 41.28 & Atypical 2 & Ambig.$^{**}$ & ${24.38}_{-0.08}^{+0.07}$* & 43.21 & Sab & pair \\
Mrk\,266SW & 13h38m17.31s & 48d16m32.0s & 0.0287 & 42.45 & 42.2 & 40.98 & Atypical 2 & -- & < 20* & 44.06 & Sab & pair \\
TOLOLO\,00113 & 13h54m15.41s & -37d46m33.2s & 0.0508 & 44.21 & 43.75 & 42.79 & Atypical 2 & --  & 22.91 & 43.34 & --  & -- \\
NGC\,5643 & 14h32m40.74s & -44d10m27.9s & 0.0026 & 42.52 & 42.10 & 40.34 & Atypical 2 &  Linear$^7$ & 25.40 & 43.20 & Sc & -- \\
MCG-01-40-001 & 15h33m20.71s & -08d42m01.9s & 0.0227 & 43.56 & 43.25 & 42.58 & Atypical 2 & --  & 22.73 & 43.50 & Sb & -- \\
CGCG\,367-009 & 16h19m19.26s & 81d02m47.6s & 0.0230 & 42.85 & 43.09 & 41.54 & Atypical 2 & --  & 23.02 & 43.47 & -- & -- \\
NGC\,6232 & 16h43m20.24s & 70d37m57.1s & 0.0148 & 42.89 & 42.84 & 40.95 & Atypical 2 & -- & 24.94 & 42.72 & Sa & group \\
LEDA\,214543 & 16h50m42.73s & 04d36m18.0s & 0.0322 & 43.20 & 43.09 & 41.82 & Atypical 2 & -- & 22.58 & 42.21 & E & -- \\
J21090996-0940147 & 21h09m09.97s & -09d40m14.7s & 0.0265 & 43.74 & 43.21 & 41.66 & Atypical 2 & -- & 21.20 & 43.40 & S0 & -- \\
J21140128+8204483 & 21h14m01.18s & 82d04m48.3s & 0.084 & 44.48 & 44.33 & 43.05 & Atypical 2 & Linear$^8$  & ${23.56}_{-0.39}^{+0.15}$* & 44.70 & --  & -- \\
\hline
NGC\,253$^{\dagger}$ & 00h47m33.12s & -25d17m17.6s & 0.0008 &  41.59 & 39.26 & 37.23 & Atypical 3 & Diffuse$^2$ & < 20*   & 46.53 & SABc & group \\
NGC\,3628$^{\dagger}$ & 11h20m16.97s & 13d35m22.9s & 0.0028 &  43.65 & 38.67 & 36.22 & Atypical 3 &  Diffuse$^2$ & ${23.38}_{-1.22}^{+0.26}$* & 42.63 & SBb & group\\
ESO\,137-G034$^{\dagger}$ & 16h35m14.11s & -58d04m48.1s & 0.0077 &  46.42 & 42.54 & 42.77 & Atypical 3 & --& 24.30 & 43.22 & SABa & group \\
ESO\,234-G050$^{\dagger}$ & 20h35m57.88s & -50d11m32.1s & 0.0087 &  46.14 & 41.62 & 40.53 & Atypical 3 & --& 23.08 & 42.39 & E & -- \\
ESO\,234-IG063$^{\dagger}$ & 20h40m15.74s & -51d25m47.1s & 0.05395 &  47.50 & 43.28 & 42.37 & Atypical 3 & -- & 23.41 & 43.91 & -- & multiple \\
\enddata
\tablecomments{\scriptsize Columns as in Table\ref{table:candidates}. References: 1-\cite{Morganti1993}; 2-\cite{Condon(1987)};3-\cite{Burns1983};4-\cite{Schmitt2001};5-\cite{Ulvestad1984};6-\cite{Ulvestad1981};7-\cite{Morris1985};8-\cite{Lara2001}. **the source NGC\,5135 is classified as Ambiguous in \cite{Ulvestad1989}; indeed, from the image we could not identify a clear sign of \textit{Linear} or \textit{Diffuse} feature.}
\end{deluxetable*}

\clearpage
\appendix

\section{\emph{NuSTAR} archival observations} \label{sec:appendixA}

We find in the sample, that 35 objects do not present a reliable obscuration measurement (those marked with an asterisk or hyphen in Col.10 in Tables \ref{table:candidates} and \ref{table:rejected}). Out of the 35, we searched in the \textit{NuSTAR} archive and found observations for 18 objects: PKS\,0558-504, NGC\,253, IRAS\,09149-6206, Mrk\,704, M\,81, ESO\,438-G009, Mrk\,1392, Mrk\,915, NGC\,4051, NGC\,3982, ESO\,509-G038, J21140128+8204483, Mrk\,1310, Mrk\,205, Mrk\,266SW, NGC\,5135, NGC\,3627, and NGC\,3628. We extracted the spectra using standard procedures, by using the analysis software \emph{NuSTARDAS} v.1.4.4, with a 60 arcsec extraction radius in all cases.\\ 
We fit the spectra with a single power-law with partial covering absorber, accounting for Galactic absorption as well. We find significant absorption in all but three sources (for which column (4) in Tab. \ref{table:nh} is marked with hyphen). In the case of NGC\,252, Mrk\,704, and NGC\,5135, we also add the Fe K$\alpha$ line at 6.4 keV. In table \ref{table:nh} we show the values found for $\rm{N_H}$, $\rm{\Gamma}$ and intrinsic luminosity in the 3-10 keV band, in all cases.

\begin{table}
\scriptsize
\centering
\begin{tabular}{lllll}
\hline
Name & $\rm{\chi^2/dof}$ &  $\rm{\Gamma}$ & $\rm{\log \ N_H }$ &$\rm{\log \ L_{(2-10])\, keV}  }$ \\ 
& & & $\rm{[cm^{-2}]}$ & $\rm{[erg \ s^{-1}]}$ \\
(1) & (2) & (3) & (4) & (5) \\\hline
PKS0558-504* & $220.99/270$ & ${2.16}_{-0.02}^{+0.03}$ & ${25.46}_{-0.50}^{+1.00}$ & ${44.76}_{-0.19}^{+0.19}$ \\ 
NGC253 & $769.88/449$ & ${2.44}_{-0.02}^{+0.01}$ & $-$ & ${39.58}_{-0.01}^{+0.01}$ \\ 
IRAS09149-6206 & $1324.1/1015$ & ${1.95}_{-0.02}^{+0.02}$ & ${24.19}_{-0.05}^{+0.05}$ & ${44.28}_{-0.01}^{+0.01}$ \\ 
Mrk704 & $400.8/342$ & ${1.49}_{-0.03}^{+0.05}$ & ${22.72}_{-0.07}^{+0.51}$ & ${43.32}_{-0.11}^{+0.11}$ \\ 
M81 & $1422.1/1323$ & ${2.11}_{-0.02}^{+0.02}$ & ${23.58}_{-0.03}^{+0.03}$ & ${40.693}_{-0.003}^{+0.003}$ \\ 
ESO438-G009 & $105.6/137$ & ${1.84}_{-0.06}^{+0.08}$ & ${24.51}_{-0.31}^{+0.46}$ & ${42.78}_{-0.04}^{+0.04}$ \\ 
Mrk1392 & $220.71/198$ & ${1.84}_{-0.04}^{+0.05}$ & ${24.63}_{-0.18}^{+0.16}$ & ${43.36}_{-0.03}^{+0.03}$ \\ 
Mrk915 & $470.71/493$ & ${1.87}_{-0.06}^{+0.06}$ & ${23.53}_{-0.09}^{+0.07}$ & ${43.11}_{-0.02}^{+0.02}$ \\ 
NGC4051 & $2177.02/1619$ & ${1.85}_{-0.01}^{+0.01}$ & ${24.53}_{-0.02}^{+0.02}$ & ${43.563}_{-0.003}^{+0.003}$ \\ 
NGC3982 & $51.25/42$ & ${3.06}_{-0.76}^{+1.27}$ & ${23.83}_{-0.20}^{+0.18}$ & ${41.12}_{-1.06}^{+1.06}$ \\ 
ESO509-G038 & $250.34/243$ & ${2.12}_{-0.11}^{+0.10}$ & ${23.90}_{-0.06}^{+0.07}$ & ${43.34}_{-0.05}^{+0.05}$ \\ 
J21140128+8204483 & $188.20/196$ & $1.96_{-0.11}^{+0.11}$ & ${23.56}_{-0.39}^{+0.15}$ & $44.49_{-0.07}^{0.07}$ \\ 
Mrk1310 & $300.03/285$ & ${1.77}_{-0.02}^{+0.02}$ & $-$ & ${42.89}_{-0.01}^{+0.01}$ \\ 
Mrk205 & $278.9/297$ & ${2.14}_{-0.10}^{+0.10}$ & ${23.69}_{-0.13}^{+0.09}$ & ${44.24}_{-0.04}^{+0.04}$ \\ 
Mrk266SW & $51.6/52$ & ${1.08}_{-0.15}^{+0.15}$ & $-$ & ${41.84}_{-0.09}^{+0.09}$ \\ 
NGC5135 & $77.8/52$ & ${1.25}_{-0.18}^{+0.17}$ & ${24.38}_{-0.08}^{+0.07}$ & ${39.98}_{-0.13}^{+0.13}$ \\ 
NGC3627 & $60.59/67$ & ${2.57}_{-0.21}^{+0.45}$ & ${24.26}_{-0.45}^{+0.67}$ & ${<40.20}$ \\ 
NGC3628 & $65.58/69$ & ${2.44}_{-0.24}^{+0.29}$ & ${23.38}_{-1.22}^{+0.26}$ & ${40.14}_{-0.41}^{+0.41}$ \\ 
M51 & $170.35/166$ & ${1.78}_{-0.09}^{+0.09}$ & ${24.67}_{-0.06}^{+0.06}$ & ${40.23}_{-0.04}^{+0.04}$ \\ 
       \hline     \hline
    \end{tabular}
    \caption{Spectral fit for those objects with available \textit{NuSTAR} observations. Colum (1) is the object name, column (2) are the statistics, $\rm{\chi^2/dof}$, column (3) is the photon index, column (4) is the column density in units of $\rm{cm^{-2}}$, and column (5) is the intrinsic luminosity in the 2-10 keV band.}
    \label{table:nh}
\end{table}

\section{Multi-wavelength diagnostic}

\subsection{Optical diagnostic}
\label{app:bpt_diagram}

To create the Baldwin, Phillips \& Telervich (BPT) diagrams \citet{Baldwin81}, we compile the reddened corrected fluxes available measurements of the [O\,III] $5007$ \AA, [N\,II] $6583$ \AA, H$\beta$ and H$\alpha$ emission lines presented in \emph{Swift} BAT 70-month catalog \citep[][]{Koss17}. In Tables\,\ref{tab:fluxes} and \ref{tab:fluxes2} we include these optical emission line fluxes for the accepted fading/rising and atypical candidates, respectively.

\subsection{Infrared diagnostic}
\label{app:SpectralFit}

We convert IRS/\emph{Spitzer} spectra into X-ray spectral fitting package XSPEC \citep{Arnaud96} format using {\sc flx2xsp} task within HEASOFT\footnote{https://heasarc.gsfc.nasa.gov}. Following the technique developed by \citet{Gonzalez-Martin09B}, we fit each spectrum using four baseline models:
    \begin{eqnarray}
        M_1 & = & zdust \times {AGN \,dust} \label{eqs:1} \\
        M_2 & = & zdust \times {AGN \,dust} + Stellar \label{eqs:2} \\
        M_3 & = & zdust \times  {AGN \,dust} + ISM \label{eqs:3} \\
        M_4 & = & zdust \times {AGN \,dust} + ISM + Stellar \label{eqs:4}
    \end{eqnarray}

\noindent where $zdust$ component is the foreground extinction by dust grains \citep{Pei92}. The $ISM$ and $stellar$ components consider circumnuclear components such as the interstellar medium and stellar population of $\rm{10^{10}}$ years and solar metallicity. These components were taken from \citet{Smith07} and \citet{Bruzual03}, respectively. Finally, the $AGN \, dust$ corresponds to the smooth by \citet{Fritz06} [F06], clumpy by \citet{Nenkova08B} [N08] and \citet{Hoenig10} [H10], the two phase (clumpy and smooth) by \citet{Stalevski16} [S16] torus models, and the clumpy disk-wind model by \citet{Hoenig17} [H17] designed to describe the IR AGN emission. We compute the $\chi^2$ statistics throughout the analysis to find the absolute minimum for each parameter as the best fit. We then use f-statistics to test whether the inclusion of the stellar (eq.\,\ref{eqs:2}), ISM (eq.\,\ref{eqs:3}), and/or the stellar+ISM (eq.\,\ref{eqs:4}) components significantly improves the simpler model when f-test probability is below $10^{-4}$. If several models describe equally well the data (according to the $\rm{\chi^2}$ statistics and using the Akaike criteria \citep[see Eqs.\,5-7 by][]{Emmanoulopoulos16}) we include both models as feasible representation of the data. We provide good spectral fitting ($\rm{\chi^2/dof < 1.3}$ for all but one object (Mrk\,1210). 

The range that covers the \emph{Spitzer} spectra allows us to measure the [OIV] line. Therefore, we compiled the [OIV] fluxes at 25.9$\rm{\mu m}$ for the 38 objects observed with \emph{Spitzer} and available at the  Combined Atlas of Sources with \emph{Spitzer} IRS Spectra (CASSIS\footnote{https://cassis.sirtf.com}). We compute the emission-line flux by fitting a Gaussian above the continuum. Tables\,\ref{tab:fluxes} and \ref{tab:fluxes2} include the [OIV] luminosities when available for accepted and atypical candidates, respectively. Among them, two, 19, 10, one, and six are early fading, fading, atypical-2, atypical-3, and atypical-1, respectively.

\section{Radio morphology} \label{sec:RadioImages}

We search the radio images available in the literature of the 102 sources considered in Section \ref{sec:candidates}. Among the 61 bona fide candidates, we found radio images for 46 targets, while for the remaining 15, no radio images were available. Among the 41 atypical candidates, we found only 15 radio images. All the radio data we collected have been obtained using the Jansky Very Large Array (JVLA) at different frequencies from low-frequency L band (1.4 GHz) to high-frequency K band (22 GHz) and using different configurations of the interferometer array. Therefore, the angular resolution of the images ranges in a broad interval from tens to one arcseconds. We are aware that this huge range of available angular resolutions of the different observations could result in an uncertain morphological classification (e.g., extended emission that could be resolved out at higher angular resolution); new radio observations with equal observational set up for all the sources would be appropriate to eliminate this uncertainty. Therefore, we suggest taking this classification as a first hint. The classification we adopted is described as follows:

\begin{itemize}
    \item Compact source: a point-like object with unresolved radio emission;
    \item Linear source: an object showing elongated, jet-like feature;
    \item Diffuse source: an object showing extended, rounded emission feature.
\end{itemize}

\begin{figure*}
\begin{center}
\includegraphics[width=1.0\columnwidth]{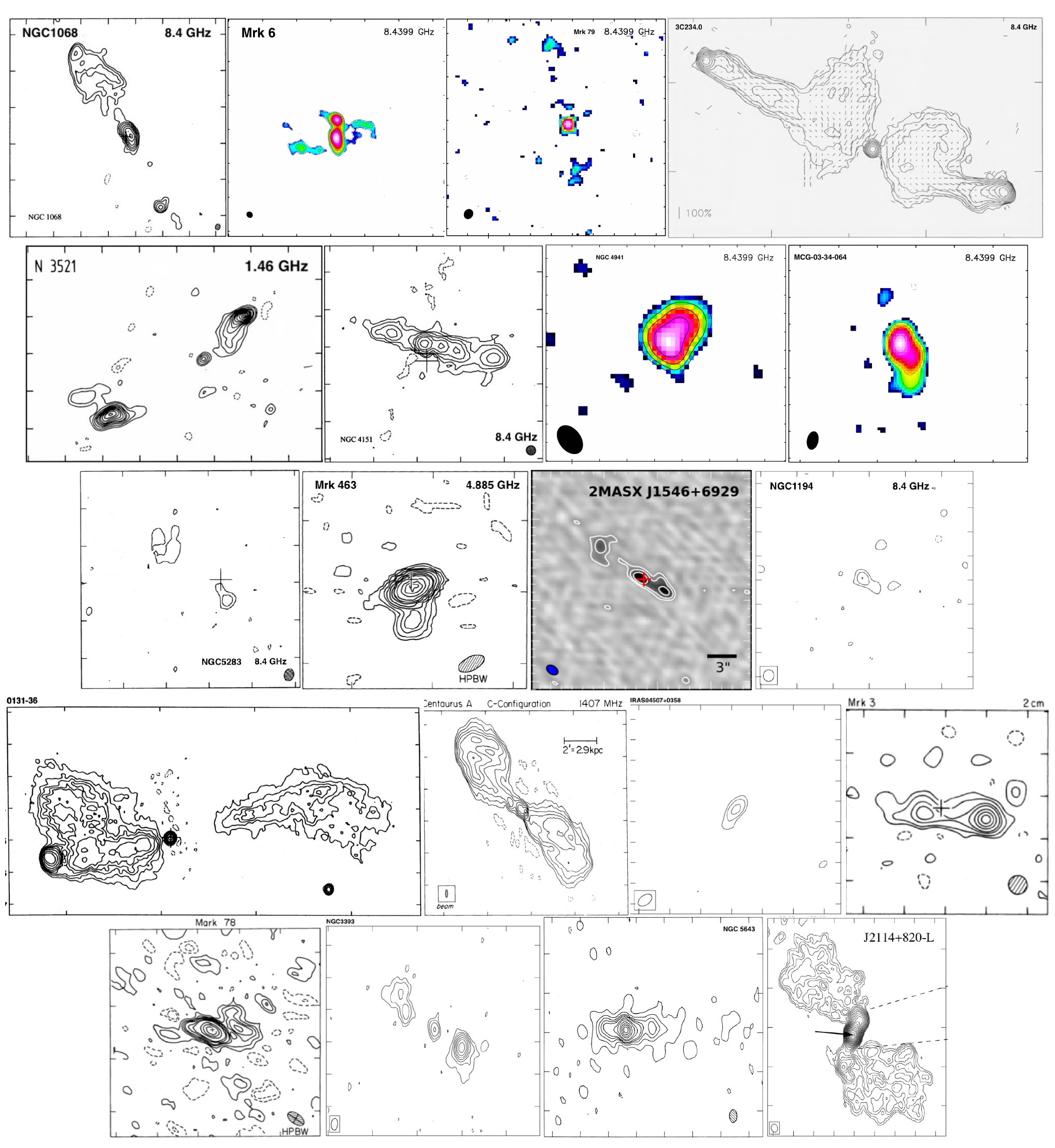}
\caption{Images of the objects classified as \textit{Linear} among the bona fide and atypical sources. All the data are JVLA images at different angular resolutions; references are reported at Table \ref{table:candidates}.}
\label{LINEAR}
\end{center}
\end{figure*}

\begin{figure*}
\begin{center}
\includegraphics[width=1.0\columnwidth]{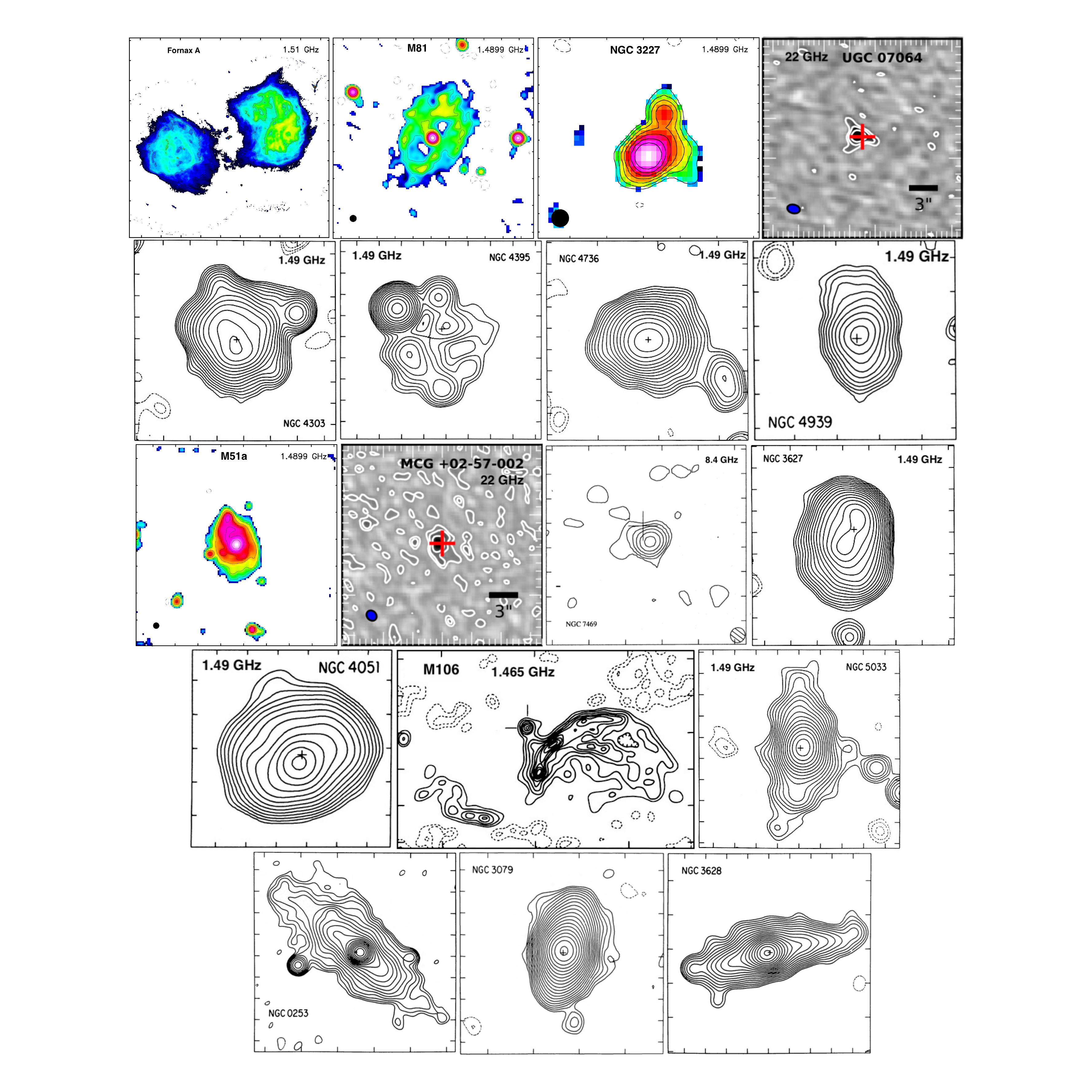}
\caption{Images of the objects classified as \textit{Diffuse} among the bona fide and atypical sources. All the data are JVLA images at different angular resolutions; references are reported at Table \ref{table:rejected}.}
\label{EXTENDED}
\end{center}
\end{figure*}

The morphological classification together with the relative references for the \textit{Linear} and \textit{Diffuse} images are available in Col.\,9 at Tables\,\ref{table:candidates} and \ref{table:rejected}.
The images found in the literature for the \textit{Linear} and \textit{Diffuse} sources are collected in Figs.\,\ref{LINEAR} and \ref{EXTENDED}.

\begin{deluxetable}{clcccc|ccccccc}
\tablecolumns{13}
\tablewidth{10cm}
\tablecaption{Fluxes for fading/rising candidates (same as Table\,\ref{table:candidates}).\label{tab:fluxes}}
\tabletypesize{\tiny}
\tablehead{
    \\
 & \multicolumn4c{Optical} & \multicolumn5c{Mid-infrared} & Classification \\ 
\colhead{Name} & \colhead{$\rm{F_{H\beta}}$} &  \colhead{$\rm{F_{[O\,III]}}$} & \colhead{$\rm{F_{H\alpha}}$} & \colhead{$\rm{F_{[N\,II]}}$} & \colhead{$\rm{log(L_{[OIV]})}$} & \colhead{model} & \colhead{AGN/Stellar/ISM} & \colhead{$\chi^2_{r}$} & \colhead{$\rm{E_(B-V)}$} & \\
& \multicolumn4c{[$10^{-15}$\,W\,m$^{-2}$]} & \colhead{[$\rm{erg s^{-1}}$]} &  & \colhead{[$\%$]} &  &  & \\
\colhead{(1)} & \colhead{(2)} &  \colhead{(3)} & \colhead{(4)} & \colhead{(5)} & \colhead{(6)} & \colhead{(7)} & \colhead{(8)} & \colhead{(9)} & \colhead{(10)} & \colhead{(11)}
}
\startdata
Mrk\,335 & $82\pm15.7$ & $276.2\pm6.5$ & $243.4\pm6.9$ & $22.9\pm2.6$ & $-$ & $-$ & $-$ & $-$ & $-$ & S1.2 \\
J00430184+3017195 & $25.8$ & $91.5\pm2.7$ & $-$ & $-$  & $-$ & $-$ & $-$ & $-$ & $-$ &  S2 \\
Mrk\,359 & $22.2\pm1$ & $122.4\pm0.1$ & $108.6\pm2.2$ & $45\pm1$ & $-$ & $-$ & $-$ & $-$ & $-$ & S1.5  \\
NGC\,1068 & $2208.7\pm13.5$ & $27677.7\pm32.8$ & $6497.5\pm3.7$ & $11935\pm8$ & $-$ & $-$ & $-$ & $-$ & $-$ & S2 \\
Fornax\,A & $-$ & $-$ & $-$ & $-$ & $-$ & $-$ & $-$ & $-$ & $-$ & L2 \\
IRAS\,04124-0803 & $32.9\pm10.4$ & $441.8\pm3.3$ & $369.7\pm0.8$ & $45.8\pm0.9$ & $-$ & $-$ & $-$ & $-$ & $-$ & S1 \\
Mrk\,618 & $37.1\pm5.8$ & $195.4\pm3.7$ & $117.1\pm7.2$ & $40.9\pm2$  & $-$ & $-$ & $-$ & $-$ & $-$ & S1 \\
LEDA\,097068 & $22.2\pm17.5$ & $392.7\pm17$ & $655.1\pm24.1$ & $604.1\pm19.2$ & $-$ & $-$ & $-$ & $-$ & $-$ & S1  \\
IRAS\,05218-1212 & $6.5\pm0.1$ & $85.5\pm0.3$ & $39.4\pm0.8$ & $21.4\pm0.5$ &43.4 $\pm$ 1.5  & H17	&  85.8/  0.0/ 14.2 &0.36& $<$0.5 & S1 \\
Mrk\,6 & $161.5\pm11.4$ & $1482.8\pm21$ & $608.9\pm3.3$ & $354\pm3.4$ &  $-$ & $-$ & $-$ & $-$ & $-$ & S1.5 \\
Mrk\,79 & $55.2\pm4.8$ & $556.2\pm4.6$ & $185.3\pm7.4$ & $133.9\pm7.5$ & $-$ & $-$ & $-$ & $-$ & $-$ & S1.2 \\
Mrk\,10 & $13.8\pm6$ & $171.1\pm0.7$ & $62.4\pm1.3$ & $45.2\pm1.2$ & 42.6 $\pm$ 1.8  & N08  &  94.3/  5.7/  0.0  &0.31& $<$0.5 & S1.2 \\
Mrk\,1210 & $73.6\pm0.2$ & $803\pm0.3$ & $173.1\pm0.6$ & $90.8\pm0.1$ & 42.1 $\pm$ 1.0 & H17   &  81.8/  1.7/ 16.5 &1.83& $<$0.5 & S1 \\
IRAS\,09149-6206 & $48.3\pm127$ & $505.9\pm104.1$ & $26.3\pm4.2$ & $13.8$ & $-$ & $-$ & $-$ & $-$ & $-$ & S1 \\
Mrk\,704 & $37.7\pm5.7$ & $337.7\pm6.2$ & $94.4\pm16.8$ & $24.6\pm10.9$ & $-$ & $-$ & $-$ & $-$ & $-$ & S1.2 \\
M\,81 & $107.4$ & $266.8\pm6.8$ & $200\pm25.4$ & $460.6\pm19.3$ & $-$ & $-$ & $-$ & $-$ & $-$ & L1.8 \\
3C\,234.0 & $15\pm0.1$ & $160.7\pm0.1$ & $56.7\pm0.3$ & $16.8\pm0.1$ & 44.24 $\pm$ 1.3 & F06 &  92.2/  2.9/  5.0 &0.37& $<$0.5 & S1 \\
NGC\,3227 & $78.6\pm3.5$ & $932.5\pm4.1$ & $386.5\pm3.2$ & $574.3\pm6.6$ &41.7 $\pm$ 1.6 & N08 &  51.0/3.3/45.7 & 1.29 & $<$0.5 & S1.5 \\
ESO\,317-G038 & $8.3\pm0.2$ & $122.8\pm0.3$ & $91.8\pm0.6$ & $88.6\pm0.6$ & $-$ & $-$ & $-$ & $-$ & $-$ & S2 \\
NGC\,3379 & $-$ & $-$ & $-$ & $-$ & $-$ & $-$ & $-$ & $-$ & $-$ & L2 \\
NGC\,3521 & $-$ & $-$ & $-$ & $-$ & $-$ & $-$ & $-$ & $-$ & $-$ & L2 \\
ESO\,438-G009 & $39.7\pm0.9$ & $75.5\pm0.3$ & $211.5\pm2.6$ & $143.3\pm1.8$ & $-$ & $-$ & $-$ & $-$ & $-$ & S1.5 \\
NGC\,3607$^{\dagger}$ & $-$ & $-$ & $-$ & $-$ & $-$ & $-$ & $-$ & $-$ & $-$ & L2 \\
PG\,1138+222 & $27.5\pm0.3$ & $200.4\pm0.4$ & $136.9\pm4.3$ & $35.8\pm1.7$ & $-$ & $-$ & $-$ & $-$ & $-$ & S1 \\
NGC\,3982 & $-$ & $-$ & $-$ & $-$ & $-$ & $-$ & $-$ & $-$ & $-$ & S2 \\
UGC\,07064 & $12.1$ & $126.7\pm0.2$ & $59\pm0.2$ & $57.6\pm0.2$ & $-$ & $-$ & $-$ & $-$ & $-$ & S1.9 \\
NGC\,4151 & $812.2\pm302$ & $10035.9\pm66.4$ & $3067\pm25.6$ & $2270.7\pm24.4$ & 41.7 $\pm$ 1.8 & N08  &  95.9/  4.1/  0.0  &0.98 & $<$0.5 & S1.5 \\
Mrk\,766 & $61.3\pm2.4$ & $463.8\pm1.1$ & $209.6\pm1.7$ & $104.3\pm0.5$ & $-$ & $-$ & $-$ & $-$ & $-$ & S1.5 \\
NGC\,4303 & $-$ & $-$ & $-$ & $-$ & $-$ & $-$ & $-$ & $-$ & $-$ & S2 \\
NGC\,4395 & $37.7\pm0.2$ & $333.5\pm0.2$ & $149.1\pm0.2$ & $33.2\pm0.1$ &  39.3 $\pm$ 1.5 & N08  &  72.7/  4.0/ 23.4 &0.69& $<$0.5 & S1.8 \\
J123212.3-421745 & $11.3\pm7.5$ & $216.2\pm11.9$ & $82.7\pm0.6$ & $2.8\pm0.9$ & $-$ & $-$ & $-$ & $-$ & $-$ & S1.5 \\
LEDA\,170194 & $65.7\pm0.1$ & $568\pm0.1$ & $611.6\pm0.1$ & $593.9\pm0.2$ & $-$ & $-$ & $-$ & $-$ & $-$ -& S2 \\
NGC\,4736 & $-$ & $-$ & $-$ & $-$ & $-$ & $-$ & $-$ & $-$ & $-$  & S2 \\
NGC\,4748 & $70.1\pm1.2$ & $531.3\pm1.1$ & $297.1\pm1.2$ & $235.5\pm1.1$ & $-$ & $-$ & $-$ & $-$ & $-$ & S1 \\
NGC\,4941 & $29\pm0.2$ & $294.7\pm0.4$ & $119.2\pm0.6$ & $194.8\pm0.6$ & 40.9 $\pm$ 1.5 & N08  & 100.0/  0.0/  0.0 &0.44& 0.7$_{0.6}^{0.8}$ & S2 \\
NGC\,4939 & $32.5\pm0.3$ & $479.5\pm0.9$ & $148.8\pm0.6$ & $226.6\pm0.9$ & 42.3 $\pm$ 1.8 & N08  &  90.7/  3.3/  6.0  &0.85& $<$0.5 & S2 \\
MCG-03-34-064 & $65.4\pm0.5$ & $996.5\pm1.5$ & $129.2\pm0.7$ & $231.3\pm0.9$ & 43.3 $\pm$ 1.6 & $-$ & circumnuclear contribution & $-$ & $-$ & S1.8 \\
M\,51a & $-$ & $-$ & $-$ & $-$ & $-$ & $-$ & $-$ & $-$ & $-$ &   S2 \\
ESO\,509-G038 & $13.7\pm0.6$ & $195.1\pm0.5$ & $49.8\pm0.8$ & $67.9\pm1$ & $-$ & $-$ & $-$ & $-$ & $-$ & S1 \\
NGC\,5283 & $40.5\pm0.8$ & $385\pm0.7$ & $139.4\pm0.1$ & $131.9\pm0.2$ & $-$ & $-$ & $-$ & $-$ & $-$ & S2 \\
NGC\,5273 & $5.9\pm0.2$ & $101.9\pm0.4$ & $27.1\pm0.1$ & $37.9\pm0.3$ & 40.2 $\pm$ 0.5 & F06 & 100.0/0.0/0.0 & 0.23 & $<$0.5 & S1.9 \\
Mrk\,463 & $60.5$ & $463.5\pm0.2$ & $141.3\pm0.3$ & $63.8\pm0.1$ & 43.9 $\pm$ 1.6 & S16   &  83.3/  4.4/ 12.3 &0.70 & 0.1$_{0.0}^{0.2}$ & S2 \\
Mrk\,477 & $95.1\pm0.2$ & $953.9\pm0.3$ & $294.9\pm0.7$ & $96\pm0.4$ & 43.3 $\pm$ 1.5 & $-$ & circumnuclear contribution & $-$ & $-$ & S1 \\
IC\,4518A & $35.1\pm0.4$ & $292.6\pm0.2$ & $160.9\pm0.3$ & $124.1\pm0.2$ & 43.07 $\pm$ 1.9 & $-$ & circumnuclear contribution & $-$ & $-$ & S2 \\
Mrk\,1392 & $11.3\pm0.5$ & $210.4\pm0.3$ & $90.6\pm0.9$ & $93.8\pm0.8$ & 43.0 $\pm$ 1.73 & N08  &  91.7/  2.4/  5.9 &0.68& $<$0.5 & S1.8 \\
        &               &               &           &               &               & H17   &  77.5/  1.5/ 21.0 &0.67& $<$0.5 \\
J15462424+6929102 & $20.5\pm0.3$ & $215.9\pm0.1$ & $66\pm0.1$ & $81.7\pm0.1$ & $-$ & $-$ & $-$ & $-$ & $-$ & S1.9 \\
J16531506+2349431 & $13.7\pm0.1$ & $163.4\pm0.1$ & $53.9\pm0.2$ & $21.5\pm0.1$ & $-$ & $-$ & $-$ & $-$ & $-$ & S2 \\
Fairall\,49 & $55.1\pm0.3$ & $267.1\pm0.3$ & $334.5\pm0.7$ & $206.3\pm0.6$ & 42.96 $\pm$ 1.4  & N08  &  66.4/  6.1/ 27.5 &1.21 & $<$0.5 & S2 \\
J19373299-0613046 & $125.5\pm55.1$ & $1153.4\pm8.3$ & $479.3\pm2.3$ & $194.3\pm1.1$ & $-$ & $-$ & $-$ & $-$ & $-$ & S1 \\
MCG+02-57-002 & $57.3\pm1.3$ & $215.8\pm0.5$ & $512.6\pm1.5$ & $252.5\pm1.3$ & $-$ & $-$ & $-$ & $-$ & $-$ & S1.5 \\
Mrk\,915 & $72.3\pm0.6$ & $761.1\pm0.7$ & $263.8\pm1.9$ & $179.6e.15\pm2.5$ & 43.1 $\pm$ 1.7 & N08  &  75.1/  4.1/ 20.8  &0.48 & $<$0.5 & S1 \\
MCG+01-57-016 & $40.9\pm1.4$ & $422.3\pm1$ & $131.7\pm0.5$ & $106.4\pm0.7$ & 42.5 $\pm$ 1.3 & N08  &  62.2/  4.7/ 33.1 &1.00& $<$0.5 & S1.8 \\
NGC\,7469 & $252.3\pm10.2$ & $1159.5\pm8.8$ & $920.7\pm5.1$ & $565.9\pm3.8$ & 42.5 $\pm$ 0.84 & $-$ & circumnuclear contribution & $-$ & $-$ & S1.2 \\
\hline
NGC\,3627 & $-$ & $-$ & $-$ & $-$ &$-$ & $-$ & $-$ & $-$ & $-$ & S2 \\
NGC\,4051 & $49.1\pm38.3$ & $310\pm6.2$ & $448.3\pm1.7$ & $148.4\pm0.5$ & $-$ & $-$ & $-$ & $-$ & $-$ & S1.5 \\
M\,106 & $52.3\pm1.5$ & $179.2\pm0.6$ & $13.1\pm2.1$ & $108.5\pm1.4$ &40.0 $\pm$ 0.9 & N08 	&  86.1/  2.5/ 11.4  &0.25& 0.5$_{0.5}^{0.6}$ & L1.9 \\
NGC\,5033 & $8.1\pm0.2$ & $38.8\pm0.3$ & $48.9\pm0.6$ & $78.8\pm1.1$ &  $-$ & $-$ & $-$ & $-$ & $-$ & S1.9 \\
NGC\,7130 & $30.3\pm0.3$ & $155.5\pm0.5$ & $119.7\pm1$ & $114\pm0.7$ & 42.2 $\pm$ 0.8 & $-$ & circumnuclear contribution & $-$ & $-$ & L1.9 \\
\hline
NGC\,1194 & $1.9$ & $29.5\pm0.3$ & $16.6\pm0.1$ & $10.3\pm0.1$ &  $-$ & $-$ & $-$ & $-$ & $-$ & S1.9 \\
J14391186+1415215 & $0.5$ & $0.8$ & $2$ & $1.3$ &  $-$ & $-$ & $-$ & $-$ & $-$ & S1  \\\hline
J08551746-2854218 & $3.8\pm0.2$ & $0.7\pm0.1$ & $4.8\pm0.1$ & $2.9$ &  $-$ & $-$ & $-$ & $-$ & $-$ & S2 \\
\enddata
\tablecomments{\scriptsize Column (1): Source name; Columns (2 - 5): optical fluxes obtained from \emph{Swift} BAT 70-month catalog \citep[][]{Ricci17}, which were used in the BPT diagram (Fig.\,\ref{fig:BPT}). All the fluxes are in units of $10^{-15}$\,W\,m$^{-2}$. Column (6): [OIV] luminosity; Columns (7-10) Best-fit results per object. Column (7) Models used to fit the data: F06: [Fritz06]; N08: [Nenkova08]; S16: [Stalev16]; and H17: [Hoenig17].Column (8) includes the percentage contribution to the $5-30 \, \mu m$  waveband per component (A: AGN; S: stellar; and I: ISM); Column (9) the reduced $\chi^2$ ($\chi^2/dof$); Column (10) color excess for the foreground extinction E(B-V); Column (11) is the classification retrieved from Hyperleda, NED and in a few cases, from \citet{Ichikawa17}.}
\end{deluxetable}

\begin{deluxetable}{clcccc|ccccccc}
\tabletypesize{\scriptsize}
\tablecolumns{13}
\tablewidth{10cm}
\tablecaption{Fluxes for atypical candidates (same as Table\,\ref{table:rejected}).\label{tab:fluxes2}}
\tabletypesize{\tiny}
\tablehead{
\\
 & \multicolumn4c{Optical} & \multicolumn5c{Mid-infrared} & Classification \\ 
\colhead{Name} & \colhead{$\rm{F_{H\beta}}$} &  \colhead{$\rm{F_{[O\,III]}}$} & \colhead{$\rm{F_{H\alpha}}$} & \colhead{$\rm{F_{[N\,II]}}$} & \colhead{$\rm{log(L_{[OIV]})}$} & \colhead{model} & \colhead{AGN/Stellar/ISM} & \colhead{$\chi^2_{r}$} & \colhead{$\rm{E_(B-V)}$} & \\
& \multicolumn4c{[$10^{-15}$\,W\,m$^{-2}$]} & \colhead{[$\rm{erg s^{-1}}$]} &  & \colhead{[$\%$]} &  &  & \\
\colhead{(1)} & \colhead{(2)} &  \colhead{(3)} & \colhead{(4)} & \colhead{(5)} & \colhead{(6)} & \colhead{(7)} & \colhead{(8)} & \colhead{(9)} & \colhead{(10)} & \colhead{(11)}
}
\startdata
NGC\,612$^{\dagger}$ & $3$ & $5.9\pm0.3$ & $6.5\pm0.8$ & $6.5\pm0.4$ & 42.3 $\pm$ 1.3 & $-$ & circumnuclear contribution & $-$ & $-$ & S2 \\
J02420381+0510061$^{\dagger}$ & $1.4\pm0.1$ & $1.4\pm0.3$ & $7.7$ & $4.6$ & - & - & - & - & - & S2 \\
J04440903+2813003$^{\dagger}$ & $2.1\pm0.1$ & $2.3\pm0.2$ & $10.1\pm0.1$ & $12.5\pm0.1$ & - & - & - & - & - & S2 \\
PKS\,0558-504$^{\dagger}$ & $3.5\pm1.1$ & $3.7\pm0.5$ & $41.8\pm0.4$ & $15.9\pm0.5$ & $-$ & S16 & 84.2/4.6/11.1 & 0.46 & $<$0.5 & S1 \\
NGC\,3079$^{\dagger}$ & $0.6$ & $0.8\pm0.1$ & $6.9\pm0.2$ & $8.7\pm0.2$ & 41.2 $\pm$ 1.2 & $-$ & circumnuclear contribution & $-$ & $-$ & L2 \\
Cen\,A$^{\dagger}$ & $-$ & $-$ & $-$ & $-$ & 40.6 $\pm$ 1.3 & $-$ & circumnuclear contribution & $-$ & $-$ & S2 \\ 
ESO\,097-G013$^{\dagger}$ & $-$ & $-$ & $-$ & $-$ & $-$ & $-$ & circumnuclear contribution & $-$ & $-$ & S2 \\
MCG+04-48-002$^{\dagger}$ & $0.8$ & $1.4\pm0.1$ & $23.5\pm0.1$ & $8.4\pm0.2$ & 42.0 $\pm$ 1.2 & $-$ & circumnuclear contribution & $-$ & $-$ & S2 \\
\hline
MCG-07-03-007 & $6.4\pm0.7$ & $107.7\pm6.4$ & $42.6\pm7$ & $68.6\pm8.9$ & - & - & - & - & -  & S2  \\
MCG+08-03-018 & $81.7\pm0.4$ & $1153.6\pm0.2$ & $236.6\pm0.1$ & $75.4\pm0.1$ & - & - & - & - & - & S2  \\
NGC\,526A & $54.1\pm0.2$ & $594.8\pm0.9$ & $147.7\pm0.3$ & $121.9\pm0.3$ & 42.4 $\pm$ 1.5 & H17 & 100.0/0.0/0.0 & 0.20 & 0.3$_{0.3}^{0.4}$ & S1.5 \\
  &  & & &  &  & H10 & 94.8/  5.2/  0.0 &   0.29 & 0.3$_{0.2}^{0.4}$  \\
NGC\,1229 & $13.8\pm0.1$ & $116.7\pm0.4$ & $44.1\pm0.1$ & $33.7\pm0.3$ & - & - & - & - & -   & S2 \\
J03305218+0538253 & $70.9\pm5.5$ & $587.1\pm3.6$ & $363\pm3.1$ & $27.8\pm1.6$ & - & - & - & - & - & S1 \\
CGCG\,420-015 & $33.9\pm0.2$ & $361.5\pm0.4$ & $90.9\pm0.1$ & $40.2\pm0.1$ & 42.7 $\pm$ 1.1 & H17 & 84.5/0.0/15.5 & 0.38 & 0.0$_{0.0}^{0.1}$ & S2 \\
Mrk\,3 & $600.3\pm4$ & $6853.8\pm3.7$ & $1049.2\pm2.8$ & $1401.2\pm1.8$ & 43.1 $\pm$ 1.8 & $-$ & circumnuclear contribution & $-$ & $-$ & S2 \\
Mrk\,78 & $25.6\pm0.1$ & $388.2\pm0.3$ & $185.3\pm7.4$ & $133.9\pm7.5$ & 43.7 $\pm$ 1.9 & N08 & 82.9/3.6/13.5 & 0.82 & 0.5$_{0.4}^{0.5}$ & S2 \\
J09172716-6456271 & $15.2\pm0.4$ & $131.7\pm0.1$ & $39.5\pm0.9$ & $13.6\pm0.5$ & - & - & - & - & - & S2 \\
ESO\,374-G044 & $12.2$ & $247.2\pm0.5$ & $72.1\pm0.3$ & $59.4\pm0.2$ & 42.9 $\pm$ 1.7 & H17 & 87.6/0.0/12.4 & 1.00 & $<$0.0 & S2 \\
         &  & & & & & N08 & 97.2/  2.8/  0.0 &   1.14 & $<$0.5 \\
NGC\,3393 & $291.4\pm6.9$ & $2557.2\pm11$ & $464\pm1.4$ & $491.5\pm1.6$ & 43.2 $\pm$ 2.2 & N08 & 82.3/2.0/15.8 & 0.99 & $<$0.5 & S2 \\
ESO\,265-G023 & $23.8$ & $163.3\pm0.6$ & $23\pm0.9$ & $7.4\pm0.9$ & - & - & - & - & -  & S1 \\
Mrk\,1310 & $11.6\pm0.2$ & $107.1\pm0.3$ & $46.8\pm0.5$ & $18.8\pm0.2$ & - & - & - & - & - & S1   \\
Mrk\,205 & $29.5\pm16$ & $244.2\pm12.1$ & $365.6\pm12.2$ & $177.2\pm30$ & - & - & - & - & -   & S1 \\
J12313717-4758019 & $19.1\pm0.5$ & $151.3\pm0.2$ & $74.1\pm0.4$ & $38\pm1.2$ & - & - & - & - & -  & S1 \\
NGC\,4507 & $341.2\pm1.4$ & $2902.1\pm2.6$ & $648.9\pm6.8$ & $326.5\pm2$ & 42.4 $\pm$ 1.3 &  N08 & 83.2/4.7/12.1 & 0.99 & $<$0.5  & S2 \\
ESO\,323-32 & $11.3$ & $131.2\pm0.2$ & $24.9\pm0.3$ & $55.7\pm0.6$ & - & - & - & - & - & L2 \\
Mrk\,783 & $23.9\pm5.4$ & $205.7\pm0.9$ & $192.2\pm1.6$ & $75.9\pm0.3$ & - & - & - & - & - & S1.5  \\
NGC\,5135 & $-$ & $-$ & $-$ & $-$ & 42.8 $\pm$ 1.4 & $-$ & circumnuclear contribution & $-$ & $-$ & S2 \\
Mrk\,266SW & $-$ & $-$ & $-$ & $-$ & 42.7 $\pm$ 1.8 & $-$ & circumnuclear contribution & $-$ & $-$ & L2  \\
TOLOLO\,00113 & $87\pm0.3$ & $996.1\pm1.3$ & $-$ & $-$ & - & - & - & - & - & S1.9\\
NGC\,5643 & $48.9\pm0.7$ & $635.4\pm0.6$ & $194.3\pm0.3$ & $240.9\pm0.5$ & 41.2 $\pm$ 1.3 & $-$ & circumnuclear contribution & $-$ & $-$ & S2 \\
MCG-01-40-001 & $4.7$ & $25.1\pm0.4$ & $2021.6\pm11.4$ & $1997.7\pm10.4$ & - & - & - & - & - & S2   \\
CGCG\,367-009 & $37.6\pm0.6$ & $286.3\pm1.1$ & $273.4\pm0.6$ & $277.6\pm1.2$ & - & - & - & - & - & S2 \\
NGC\,6232 & $28.5\pm0.1$ & $177.7$ & $72.9$ & $83.8\pm0.1$ & - & - & - & - & - & S2   \\
LEDA\,214543 & $22.8$ & $279.9\pm1.3$ & $265.1\pm3.6$ & $420.5\pm2.6$ & - & - & - & - & - & S2 \\
J21090996-0940147 & $36.1\pm2.3$ & $277.4\pm1.9$ & $144.2\pm27.4$ & $68.5\pm4.9$ & - & - & - & - & - & S1.2 \\
J21140128+8204483 & $31.8\pm10.7$ & $655.8\pm179.6$ & $255.1\pm5.3$ & $145.7\pm3.7$ & - & - & - & - & - & S1 \\
\hline
NGC\,253$^{\dagger}$ & $-$ & $-$ & $-$ & $-$ & - & - & - & - & -  & S2 \\
NGC\,3628$^{\dagger}$ & $-$ & $-$ & $-$ & $-$ & $-$ & $-$ & circumnuclear contribution & $-$ & $-$ & L2 \\
ESO\,137-G034$^{\dagger}$ & $3409\pm66.4$ & $41158.5\pm78.1$ & $-$ & $-$ & - & - & - & - & - & S2 \\
ESO\,234-G050$^{\dagger}$ & $61.4\pm0.3$ & $196.2\pm0.6$ & $161.3\pm0.6$ & $36.8\pm0.3$ & - & - & - & - & - & S2  \\
ESO\,234-IG063$^{\dagger}$ & $30.6\pm0.2$ & $336.1\pm0.7$ & $72.5\pm0.3$ & $30\pm0.3$ & - & - & - & - & - & S2  \\
\enddata
\tablecomments{\scriptsize Columns descriptions as in Table\,\ref{tab:fluxes}.}
\end{deluxetable}

\clearpage
\bibliography{AG_N_ing.bbl}{}
\bibliographystyle{aasjournal}



\end{document}